\documentclass[reprint,superscriptaddress]{revtex4-2} 
\usepackage{amsfonts}
\usepackage{amssymb}
\usepackage{amsmath}
\usepackage{upgreek}
\usepackage{bm}
\usepackage{physics}
\usepackage{graphicx}
\usepackage{gensymb,color}
\usepackage{xr}

\def\eqS{\begin{equation}}
\def\eqE{\end{equation}}

\makeatletter
\def\@fnsymbol#1{\ensuremath{\ifcase#1 \or \dagger\or *\or \mathsection\or \mathparagraph\or \|\or **\or \dagger\dagger \or \ddagger\ddagger \else\@ctrerr\fi}}
\makeatother

\makeatletter
\newcommand*{\balancecolsandclearpage}{%
  \close@column@grid
  \cleardoublepage
  \twocolumngrid
}

\makeatother

\begin{document}

\title{Modular connectivity in neural networks emerges from Poisson noise-motivated regularisation, and promotes robustness and compositional generalisation}

\author{Daoyuan Qian}
\affiliation{McGovern Institute for Brain Research, Massachusetts Institute of Technology, MA 02139, U.S.A.}
\affiliation{K. Lisa Yang Integrative Computational Neuroscience Center in the Yang-Tan Collective}
\affiliation{Department of Brain and Cognitive Sciences, Massachusetts Institute of Technology, MA 02139, U.S.A.}
\author{Qiyao Liang}
\affiliation{McGovern Institute for Brain Research, Massachusetts Institute of Technology, MA 02139, U.S.A.}
\affiliation{K. Lisa Yang Integrative Computational Neuroscience Center in the Yang-Tan Collective}
\affiliation{Department of Electrical Engineering and Computer Science, Massachusetts Institute of Technology, MA 02139, U.S.A.}
\author{Ila Fiete}
\thanks{fiete@mit.edu}
\affiliation{McGovern Institute for Brain Research, Massachusetts Institute of Technology, MA 02139, U.S.A.}
\affiliation{K. Lisa Yang Integrative Computational Neuroscience Center in the Yang-Tan Collective}
\affiliation{Department of Brain and Cognitive Sciences, Massachusetts Institute of Technology, MA 02139, U.S.A.}

\date{\today}

\begin{abstract}
Circuits in the brain commonly exhibit modular architectures that factorise complex tasks, resulting in the ability to compositionally generalise and reduce catastrophic forgetting. In contrast, artificial neural networks (ANNs) appear to mix all processing, because modular solutions are difficult to find as they are vanishing subspaces in the space of possible solutions. Here, we draw inspiration from fault-tolerant computation and the Poisson-like firing of real neurons to show that activity-dependent neural noise, combined with nonlinear neural responses, drives the emergence of solutions that reflect an accurate understanding of modular tasks, corresponding to acquisition of a correct world model. We find that noise-driven modularisation can be recapitulated by a deterministic regulariser that multiplicatively combines weights and activations, revealing rich phenomenology not captured in linear networks or by standard regularisation methods. Though the emergence of modular structure requires sufficiently many training samples (exponential in the number of modular task dimensions), we show that pre-modularised ANNs exhibit superior noise-robustness and the ability to generalise and extrapolate well beyond training data, compared to ANNs without such inductive biases. Together, our work demonstrates a regulariser and architectures that could encourage modularity emergence to yield functional benefits.
\end{abstract}

\maketitle

\section{Introduction}

The crux of biological organization is modularity: multiple specialist structures that operate on subsets of inputs, and are combined in various ways to enable adaptive responses to highly diverse situations. Evolutionary theorists have proposed various reasons for the emergence of modular structure, including the need for rapid adaptation to non-stationary conditions, robustness, and evolvability \cite{lorenzEmergenceModularityBiological2011}. In the context of brains and intelligence, modular systems hypothetically enable the generation of `infinite use from finite means' \cite{chomskyASPECTSTHEORYSYNTAX}, allow for learning sample efficiency \cite{boopathyBREAKINGNEURALNETWORK2025,veniatEFFICIENTCONTINUALLEARNING2021,aletNeuralRelationalInference,parascandoloLearningIndependentCausal2018}, avoid catastrophic forgetting by reusing rather than slowly re-learning and overwriting computational primitives \cite{ellefsenNeuralModularityHelps2015, hamidiModularGrowthHierarchical2024}, and enable fault tolerance \cite{mccourtNoisyDynamicalSystems2023}. Feedforward neural networks trained on multiple tasks were also found to exhibit increased modularity in tandem with performance increases \cite{guEmergenceReconfigurationModular2024}. Intuitively, a modular network structure allows disentangled processing to occur with minimum interference between unrelated information channels. However, modern machine learning methods, including stochastic gradient descent applied to large neural networks, tend not to find modular solutions even for inherently modular tasks. Furthermore, when supplied with modular representations in the form of independent task factors, ANNs usually do not then learn to compose them \cite{monteroROLEDISENTANGLEMENTGENERALISATION2021, liangCompositionalGeneralizationRequires2025, liangHowDiffusionModels2024}.
\\

The functional benefits of modular networks call for efforts to discover possible drivers of modularisation and to find the best ways to incorporate these modularity emergence mechanisms in ANNs. In particular, how might an ANN automatically identify the modular structure of a task and re-organise its connections accordingly? In a study involving boolean networks that implement digital functions, it was found that they evolve towards the vanishingly small space (in the space of all solutions) of modular solutions when they are forced to perform fault-tolerant computation in the presence of bit-flip noise \cite{mccourtNoisyDynamicalSystems2023}. Separately, it was shown that penalising large neuron activation values under a non-negative activation constraint, a form of activity regularisation, can promote the emergence of factorised representations \cite{whittingtonDisentanglementBiologicalConstraints2023}. Can noise-based modularisation from the former work be similarly induced in ANNs, using neuron noise? And, can the observation that noise is related to regularisation \cite{camutoExplicitRegularisationGaussian,pooleAnalyzingNoiseAutoencoders2014} be leveraged to define a new regulariser that serves as a generic bias towards the emergence of modular ANN solutions?
\\


Here we investigate the potential of neural noise to drive modularity emergence in ANNs over a range of noise characteristics, and explore how modularity emergence can affect the noise-robustness and generalisation ability of networks in downstream tasks. Neural noises are typically modelled by additive Gaussian noise in artificial networks. Characterisation of biological neurons, however, showed that the variance in firing rate changes with the mean firing rate, consistent with Poisson statistics in some cases \cite{softkyHighlyIrregularFiring1993} and more generally can be fitted to a power law \cite{gurResponseVariabilityNeurons1997}. Motivated by these observations, we consider noises that scale with neuron activation in ANNs. Denoting the activity output of a single neuron as $h$, and the corresponding noise as $\eta$, we draw noise values from a zero-mean Gaussian distribution with a variance dependent on $h$ with an exponent $\alpha$ as $\expval{\eta^{2}}_{n}\propto|h|^{\alpha}$, where $\expval{\dots}_{n}$ denotes averaging over noise realisations (Fig.~\ref{fig:setup}a). We find that networks trained on modular tasks find non-modular solutions when the neurons are noise-free or the noise is additive ($\alpha=0$). Modular connectivity emerges in nonlinear networks for $\alpha>0$, for instance when the noise is Poisson ($\alpha=1$) or multiplicative ($\alpha=2$). Next, we derive how the modularisation effects of training with stochastic neural activations can be recapitulated by a specific non-stochastic regularisation in the loss function. This regulariser, which we call the weighted-activity (WA) regulariser, is the product of the outgoing squared weight norm with the pre-synaptic neural activation norm. We demonstrate empirically that WA regularisation drives the emergence of input-selective neural responses that reduce overall activation (similar to  \cite{whittingtonDisentanglementBiologicalConstraints2023}), but further, it drives the network to modular solutions similar to the effects of multiplicative noise injection. We characterise the WA regulariser under different neuron activation functions, task traits, and training data density. We show that Rectified Linear Unit (ReLU) activation is important in promoting regularisation through its non-saturating behaviour, in contrast to Tanh and Sigmoid activations. When the input factors are entangled, the regulariser also finds the optimal way of dis-entangling them in the appropriate depth of a deep nonlinear network. We find that modularisation requires sufficient training data, meaning that the WA regulariser does not solve the problem of sample efficiency in module discovery. Perhaps consistent with this finding, modular organisation of the brain in nature often occurs at early stages of development, guided through processes acquired over evolutionary time-scales, and the functional benefits are only manifest when (re-)training the already modularised network on new tasks \cite{chenDevelopmentModularityNeural2015}. Motivated by this, we explore the functional advantage of modularised networks. Starting from a modular network, we show that modularity acts as a universal bias for compositional generalisation: the independent channels of processing enable the network to quickly learn the compositional structure of the task and allow it to perform well on out-of-distribution data and improving data efficiency to convergence.
\\

Taken together, our work shows that modularity in ANNs can readily emerge from ReLU activation together with activity-dependent noise injection or an equivalent WA regularisation. The latter also leads to noise-robustness, despite a deterministic training loss. As an architectural constraint, network modularisation promotes compositional generalisation. These insights may help explain the modular organisation and competence of biological neural networks and promote greater modularity in artificial ones.
\\

\section{Modular solutions emerge with non-additive neural noise}

\subsection{Network and task setup} 

\begin{figure*}[htb]
\includegraphics[width=17.8cm]{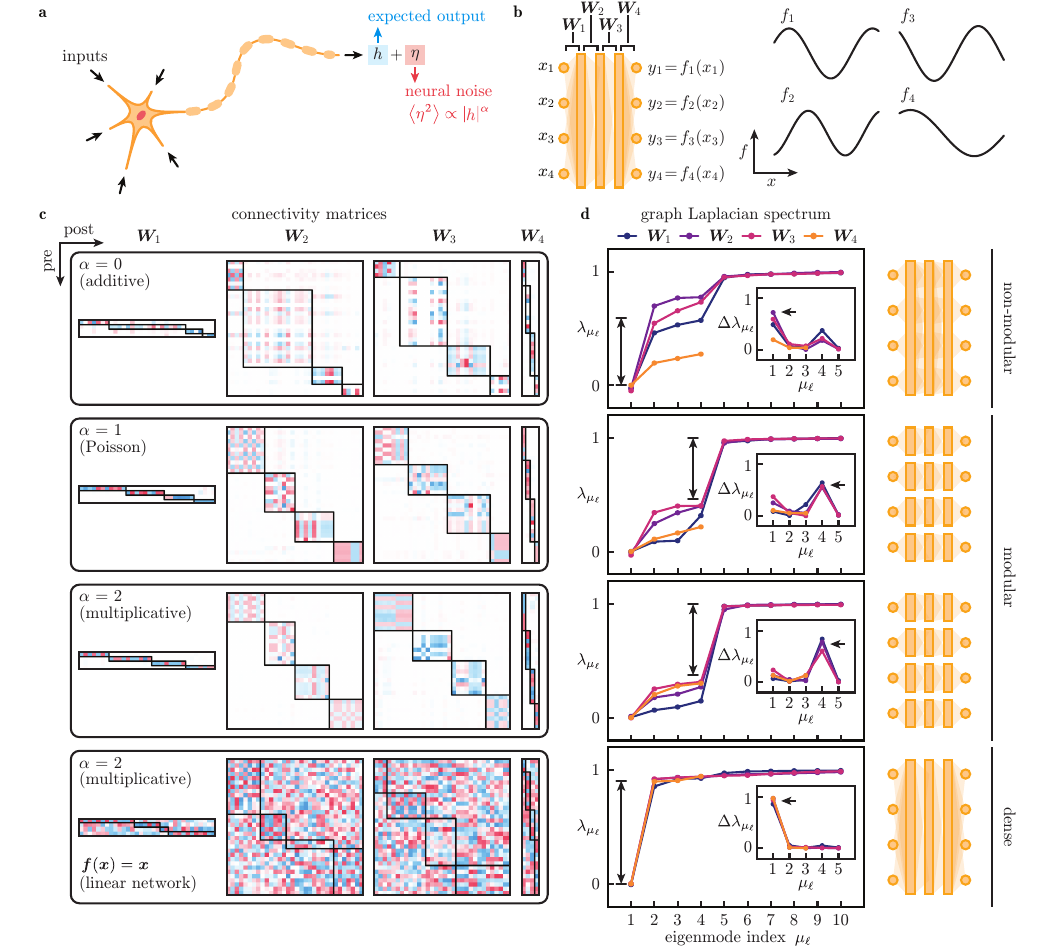}
\caption{\textbf{Non-additive noise drives deep nonlinear  ANNs to  modularise.} \textbf{a} A single neuron takes in multiple inputs and produces an output $h$ with neural noise $\eta$, the size of which scales as some power of $|h|$. \textbf{b} Modular task: a network with three hidden layers ($32$ units per intermediate layer) must learn an input-output mapping involving four independent nonlinear functions of the inputs. \textbf{c} Learned connectivity matrices across layers, for nonlinear networks and $\alpha=0$, 1, and 2 (top three rows). The network learns modular connectivity if $\alpha>0$. Rows and columns of the matrices are re-arranged according to detected clusters. When the task involves learning identity functions with a linear network and $\alpha=2$, the learned connectivity is non-modular (bottom row). 
\textbf{d} Network modularity can be characterised via the eigenspectra of the graph Laplacian matrices of the connectivities in \textbf{c}. The $\mu_{\ell}$-th eigenvalue of the $\ell$-th layer is denoted $\lambda_{\mu_{\ell}}$ (first 10 modes are shown). Clusters are detected as eigenvalues near 0, while eigenvalues closer to 1 correspond to trivial structures. The number of clusters is typically identified by finding the largest difference between neighbouring eigenvalues $\Delta\lambda_{\mu_{\ell}}\equiv\lambda_{\mu_{\ell}+1}-\lambda_{\mu_{\ell}}$ (double arrows and insets).}
\label{fig:setup}
\end{figure*}

Directly motivated by \cite{mccourtNoisyDynamicalSystems2023}, 
we consider the effects of noise in ANNs when training on modular problems. We first explore this question empirically, training a fully connected feed-forward network (3 hidden layers, 32 hidden units in each layer) to learn nonlinear function regression (Fig.~\ref{fig:setup}b) by gradient descent on the mean squared error (MSE). Each hidden unit computes the sum of its inputs and injects a noise before computing its output. Denoting inputs and outputs by vectors $\bm{x}=[x_{1},\dots,x_{4}]^{T}$ and $\bm{y}=[y_{1},\dots,y_{4}]^{T}$, the task is to learn the function $\bm{y}=\bm{f}(\bm{x})$ defined as
\eqS\label{eq:modulartask}
\bm{f}(\bm{x})=[f_{1}(x_{1}),f_{2}(x_{2}),f_{3}(x_{3}),f_{4}(x_{4})]^{T},
\eqE
where the $f_{i}$'s are four independent, randomly-generated nonlinear functions (Fig.~\ref{fig:setup}b, \cite{dq219HttpsGithubcomDq2192025b}). Inputs $\bm{x}$ are independently drawn from the uniform distribution over $(-2,2)$ so the task has a true modular structure. Given the column vector of neural activations $\bm{h}_{\ell}$ in layer $\ell$, inputs to the next layer are
\eqS
\bm{h}_{\ell+1}=\phi\left[\bm{W}_{\ell}\left(\bm{h}_{\ell}+\bm{\eta}_{\ell}\right)+\bm{b}_{\ell}\right],
\label{eq:forward}
\eqE
where $\bm{W}_{\ell}$ and $\bm{b}_{\ell}$ are weights and biases in layer $\ell$, $\bm{\eta}_{\ell}$ is a zero-mean noise vector, and $\phi(x)$ is a point-wise transfer function, which we take to be the leaky ReLU:  
\eqS
\phi(x)=\begin{cases}x&\quad\text{if }x>0,\\
0.1x&\quad\text{otherwise}.
\end{cases}
\eqE
The injected activation noise is zero-mean and Gaussian, with a covariance that scales with the norm of the summed inputs with exponent $\alpha$ (Fig.~\ref{fig:setup}a): 
\eqS
\expval{\bm{\eta}_{\ell}\bm{\eta}_{\ell}^{ T}}_{n}=\sigma_{n}^{2}\text{diag}(|\bm{h}_{\ell}|)^{\alpha},
\label{eq:noiseCov}
\eqE
for a noise strength $\sigma_{n}$. In practice, the noise $\bm{\eta}_{\ell}$ is generated by first sampling a vector of Gaussian random variables with standard deviation $\sigma_{n}$ (set to 0.1) and multiplied element-wise with $|\bm{h}_{\ell}|^{\alpha/2}$ (the exponent is calculated element-wise too). Setting $\alpha=0$, 1, and 2 corresponds to injecting additive, Poisson, and multiplicative noise, respectively. Simulation code is available on a GitHub repository \cite{dq219HttpsGithubcomDq2192025b}.
\\

\subsection{Detecting modular structure} 

A network that recognises the modular task structure should exhibit modular connectivity, such that each input $x_i$ projects without mixing with other input-driven responses in the hidden layers to the output representing $f_i(x_i)$, even though the network begins with mixing in all layers. However, there are many more alternative solutions to the task, in which the input-to-hidden layers mix the inputs in some arbitrary way and the hidden-to-output layers undo that transformation to demix them. After training, we determine whether the network exhibits modular connectivity structure by analyzing the connectivity matrix $\bm{W}_{\ell-1}$ projecting from layer $\ell-1$ (neurons indexed by $\mu_{\ell-1}$) to hidden layer $\ell$ (neurons indexed by $\mu_{\ell}$). We compute clusters of layer $\ell$ neurons that share similar input sources using the random walk variant of the graph Laplacian matrix, denoted $\bm{L}^{\text{rw}}_{\ell}$. Very briefly, each row of the weight matrix $\bm{W}_{\ell-1}$ represents an input weight vector to a post-synaptic neuron, and we calculate the row-wise normalised weights $\tilde{\bm{W}}_{\ell-1}$ by re-scaling the input vectors to have unit norm and taking the absolute value of the entries
\eqS
[\tilde{\bm{W}}_{\ell-1}]_{\mu_{\ell}\mu_{\ell-1}}\equiv\frac{|[\bm{W}_{\ell-1}]_{\mu_{\ell}\mu_{\ell-1}}|}{\sqrt{\sum_{\mu_{\ell-1}}|[\bm{W}_{\ell-1}]_{\mu_{\ell}\mu_{\ell-1}}|^{2}}}.
\eqE
We take the absolute value because we are interested in whether a post-synaptic neuron receives inputs from a pre-synaptic neuron regardless of the sign of the connectivity. $\tilde{\bm{W}}_{\ell-1}$ satisfies $\sum_{\mu_{\ell-1}}[\tilde{\bm{W}}_{\ell-1}]^{2}_{\mu_{\ell}\mu_{\ell-1}}=1$. We then calculate the (cosine) similarity matrix $\bm{S}_{\ell}$, the degree matrix $\bm{D}_{\ell}$, and the random walk Laplacian matrix $\bm{L}^{\text{rw}}_{\ell}$ as \cite{luxburgTutorialSpectralClustering2007} (the label $\ell$ is used since the $\bm{W}_{\ell-1}$ matrix is used to cluster neurons in the post-synaptic layer $\ell$)
\begin{subequations}
\begin{align}
\bm{S}_{\ell}&\equiv\tilde{\bm{W}}_{\ell-1}\tilde{\bm{W}}_{\ell-1}^{T},\\
\bm{D}_{\ell}&\equiv\text{diag}(\bm{S}_{\ell}\,\bm{1}),\\
\bm{L}_{\ell}^{\text{rw}}&\equiv\bm{I}-\bm{D}_{\ell}^{-1}\bm{S}_{\ell},
\end{align}
\end{subequations}
where $\bm{1}$ is a column vector of 1's (so matrix multiplication with $\bm{1}$ is a row-sum) and $\bm{I}$ is the identity matrix. The eigenvalues $\lambda_{\mu_{\ell}}$ of $\bm{L}_{\ell}^{\text{rw}}$ lie in the range $[0,2]$ with eigenvalues of 0, 1, and 2 indicating clusters of connected components, tree-like branches, and bipartite graphs respectively \cite{luxburgTutorialSpectralClustering2007}. With $4$ inputs, a modularised network should have 4 eigenvalues of $\bm{L}^{\text{rw}}_{\ell}$ near 0, with the rest near 1. 
\\

To visualise the modular connectivity structure, we reorder neuron indices $\mu_{\ell}$ by performing a k-means clustering of the 4 eigenvectors with the smallest eigenvalues, to group post-synaptic neurons into 4 sets. Plotting the network connectivity matrix reveals increasingly modular organization as $\alpha$ increases from 0 to 2 (Fig.~\ref{fig:setup}c, top 3 panels). After computing and sorting the eigenspectra $\{\lambda_{\mu_{\ell}}\}$ in increasing order, the largest spectral gap $\Delta\lambda_{\mu_{\ell}}\equiv\lambda_{\mu_{\ell}+1}-\lambda_{\mu_{\ell}}$ reflects the most drastic change in network structure across different eigenmodes. Consistent with standard network literature \cite{luxburgTutorialSpectralClustering2007}, we plot $\Delta\lambda_{\mu_{\ell}}$ versus $\mu_{\ell}$ to identify the number of network modules as $\mu_{\ell}* = \arg\max_{\mu_{\ell}} (\Delta\lambda_{\mu_{\ell}})$ (Fig.~\ref{fig:setup}d, double arrows). For a modular network of 4 clusters, a large eigenvalue difference is expected between the 5th (near 1) and 4th (near 0) eigenvalues. A pronounced 4-cluster connectivity structure is present only for $\alpha=1$ and 2, but not for $\alpha=0$ (Fig.~\ref{fig:setup}d, top 3 panels). As a control, we train a linear network without bias to learn the modular regression task of learning of 4 identity functions $\bm{y}=\bm{x}$ with $\alpha=2$. The resulting connectivities are dense without modular structure (Fig.~\ref{fig:setup}c, \ref{fig:setup}d, bottom panels). This shows the nonlinearity and the bias are likely essential for the emergence of ANN modularity.
\\

\subsection{Effect of noise strength} 

\begin{figure}[tb]
\includegraphics[width=8.7cm]{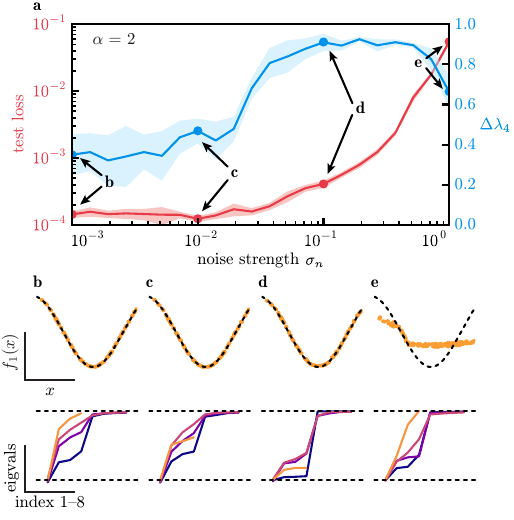}
\caption{\textbf{Noise strength leads to a loss-modularity tradeoff.} \textbf{a} In the under-parameterised small-network regime, we find that increasing the noise strength (here using multiplicative noise, $\alpha =2$) adversely affects model performance (red loss curve) while encouraging modular connectivity (blue eigen gap curve). Shaded regions are standard deviation from 5 independent runs. \textbf{b}-\textbf{e} Top: exemplar plots for model prediction (orange scatters) and ground truth (black dashed lines) of the first learned function. Bottom: spectra of the Laplacian matrices, computed in the same way as in Fig.~\ref{fig:setup}. Dashed horizontal lines are eigenvalues of 0 and 1 respectively.}
\label{fig:tradeoff}
\end{figure}

For modularisation-promoting noise, there is a tradeoff between module emergence and test loss as the noise strength $\sigma_{n}$ increases. At larger $\sigma_{n}$, the noise perturbs the gradient descent process and the test loss (mean squared error loss with noise set to 0) increases (Fig.~\ref{fig:tradeoff}a, red). Meanwhile, the modularising effect gets stronger with $\sigma_{n}$, evident from plotting the 4th eigengap $\Delta\lambda_{4}$ of the first layer ($\ell=1$) (Fig.~\ref{fig:tradeoff}a, blue). Inspecting the learned function for the first input $f_{1}(x_{1})$ and full eigenspectra at various $\sigma_{n}$ confirms the tradeoff between test loss and modularity (Fig.~\ref{fig:tradeoff}b-\ref{fig:tradeoff}e). The simulations of Fig.~\ref{fig:setup} were performed near the balance point ($\sigma_{n}=0.1$) in the two competing effects of noise, module emergence and task performance deterioration.


\section{Novel effective regularisers from noise}

To better understand why specific activity-dependent noise promotes modular connectivity, we next characterise its effects analytically. We show that the effects can be viewed as adding novel regularization terms to the loss function, allowing us to obtain non-stochastic inducers for modularity emergence. Deriving deterministic regularisers can recapitulate the beneficial effects of noise without the problems of slow learning speed and loss of performance from excess noise that are introduced by learning with noise. 

\subsection{Single-layer networks}
In a network with one hidden layer, the input ($\bm{h}_{1}=\bm{x}$), hidden ($\bm{h}_{2}$), and output ($\bm{h}_{3}=\bm{y}$) layers are related via Eq.~\eqref{eq:forward}. We follow \cite{liaoSelfAssemblyBiologicallyPlausible2024} to incorporate the effects of the nonlinearity $\phi(x)$ by defining a diagonal matrix $\bm{Q}_{\ell}$ for each hidden layer $\ell$ such that the state updates follow a (manifestly) linear form
\begin{subequations}
\begin{align}
\bm{h}_{1}&=\bm{x},\\
\bm{h}_{2}&=\bm{Q}_{2}[\bm{W}_{1}\bm{h}_{1}+\bm{b}_{1}],\\
\bm{h}_{3}&=\bm{W}_{2}(\bm{h}_{2}+\bm{\eta}_{2})+\bm{b}_{2},
\end{align}
\end{subequations}
where the $\mu_{\ell}$-th entry $[\bm{Q}_{\ell}]_{\mu_{\ell}\mu_{\ell}}$ is given by inputs from the previous layer; $\bm{Q}_{\ell}$ is different for each sample and depends on previous weights and biases. There is no $\bm{Q}_{3}$ since the output layer is linear, and there is no $\bm{\eta}_{1}$ because that would correspond to input noise. As before, the noise $\bm{\eta}$ can additive, Poisson, or multiplicative ($\alpha=\{0,1,2\}$). For $\bm{Q}_{2}$, the entries are
\eqS
[\bm{Q}_{2}]_{\mu_{2}\mu_{2}}=\begin{cases}1\quad&\text{if }[\bm{W}_{1}(\bm{h}_{1}+\bm{\eta}_{1})+\bm{b}_{1}]_{\mu_{2}}>0,\\0.1&\text{otherwise}.\end{cases}
\eqE
\\

The expected excess squared loss $\Delta L$ due to  injected noise, based on the difference in value of the network output with and without noise, is given by: 
\eqS
\begin{split}
\Delta L \equiv \expval{\Delta\bm{h}_{3}^{T}\Delta\bm{h}_{3}}_{n}&=\sigma_{n}^{2}\tr\left[\text{diag}(|\bm{h}_{2}|)^{\alpha}\bm{W}_{2}^{T}\bm{W}_{2}\right]\\
&=\sigma_{n}^{2}\sum_{\mu_{3},\mu_{2}}[\bm{W}_{2}]_{\mu_{3}\mu_{2}}^{2}|[\bm{h}_{2}]_{\mu_{2}}|^{\alpha},
\end{split}
\label{eq:MSE}
\eqE
where $\langle \cdots \rangle_n$ designates the expectation over the injected noise $\bm{\eta}$, $\mu_{3}$ and $\mu_{2}$ label the output and hidden neurons, respectively, and we have used the noise covariance Eq.~\eqref{eq:noiseCov}. This form of the excess error and the empirical results of modularity with $\alpha>0$ suggests that minimization of the products of weights and activations --- activity-dependent and weighted by outgoing connectivity --- is important. Therefore, we call these {\em weighted-activity} regularizers (WA). Below, we examine the form of the WA-$\alpha$ regularizers for different $\alpha$, neural nonlinearities, and the role of the data distribution; after that, we will examine the effects of these regularizers. 

\subsubsection{Additive noise ($\alpha=0$)}

The expression in Eq.~\eqref{eq:MSE} is not analytically tractable for general $\alpha$, but we can study special cases. For $\alpha=0$, the squared loss becomes
\eqS
\Delta L|_{\alpha=0}=\sigma_{n}^{2}\tr\left[\bm{W}_{2}^{T}\bm{W}_{2}\right] \equiv L_2.
\label{eq:MSE0}
\eqE
Thus, applying additive Gaussian noise during training is equivalent to an $L_{2}$ weight regularisation in the loss, which is in turn equivalent to weight decay in the weight updates \cite{bishopTrainingNoiseEquivalent1995}. We saw in simulations (Fig.~\ref{fig:setup}c) that additive Gaussian noise does not promote modularisation. Because $\tr\left[\bm{W}_{2}^{T}\bm{W}_{2}\right]$ is rotationally symmetric with respect to both pre and post-dynaptic neurons, mixing information channels does not induce an extra cost. This result holds for both linear and nonlinear activation functions. Thus, we surmise that even in arbitrary networks, $L_2$ weight regularisation will not effectively drive modularisation. 

\subsubsection{Multiplicative noise ($\alpha=2$), linear network}

For $\alpha=2$ the excess loss involves $\text{diag}(|\bm{h}_{2}|)^{2}$, corresponding to diagonal entries of the matrix $\bm{h}_{2}\bm{h}_{2}^{T}$. We attempt to compute this as follows: treat the network inputs  as independent Gaussian variables, with $\expval{\bm{h}_{1}\bm{h}_{1}^{T}}_{d}=\sigma_{d}^{2}\bm{I}$ where $\sigma_{d}$ is the input standard deviation and $\expval{\dots}_{d}$ designates the expectation. However, taking the expectation over $\bm{h}_{2}\bm{h}_{2}^{T}$ involves the matrices $\bm{Q}_{2}$, which depend nonlinearly on $\bm{h}_{1}$, making an explicit general expression is intractable. For a linear network, $\bm{Q}_{2}=\bm{I}$ so we obtain
\eqS
\left.\expval{\bm{h}_{2}\bm{h}_{2}^{T}}_{d}\right|_{\text{linear}}=\sigma_{d}^{2}\bm{W}_{1}\bm{W}_{1}^{T}.
\label{eq:hh}
\eqE
Because we assumed the inputs $\bm{h}_{1}$ have zero mean, the biases have dropped out in Eq.~\eqref{eq:hh}. Combining Eqs.~\eqref{eq:MSE} and \eqref{eq:hh} and writing out the indices explicitly:
\eqS
\begin{split}
\left.\expval{\Delta L}_{d}\right|_{\alpha=2\text{, linear}}&=\\
\sigma_{n}^{2}\sigma_{d}^{2}\sum_{\mu_{3},\mu_{2},\mu_{1}}[&\bm{W}_{2}]_{\mu_{3}\mu_{2}}^{2}[\bm{W}_{1}]_{\mu_{2}\mu_{1}}^{2}
\end{split}
\label{eq:lossA2Lin}
\eqE
where expectations were taken over both the noise and data. For each hidden neuron, the loss is the product of the $L_{2}$-norms of its input and output weights. From simulations (Fig.~\ref{fig:setup}c, \ref{fig:setup}d, bottom row), we know that $\alpha=2$ noise does not promote modularity in a linear network, thus the corresponding regularisation, given by the product of the squared norms of the incoming and outgoing weights at each neuron, will not lead to modularity in linear networks. 

\subsubsection{Multiplicative noise ($\alpha=2$), nonlinear network}

In a nonlinear network with $\alpha=2$, $\bm{Q}_{2}$ changes across input samples, which complicates analytical derivation. Nonetheless, we can gain some qualitative insight by approximating $\bm{h}_{1}\bm{h}_{1}^{T}\approx\sigma_{d}^{2}\bm{I}$ for each data sample. The per-sample loss of Eq.~\eqref{eq:MSE} becomes
\begin{equation}
\Delta L |_{\alpha=2}\approx\sigma_{n}^{2}\sigma_{d}^{2}\sum_{\mu_{3},\mu_{2},\mu_{1}}[\bm{W}_{2}]_{\mu_{3}\mu_{2}}^{2}[\bm{Q}_{2}]_{\mu_{2}\mu_{2}}^{2}[\bm{W}_{1}]_{\mu_{2}\mu_{1}}^{2}.
\label{eq:lossA2NonLin}
\end{equation}
Through $\bm{Q}_{2}$, we may view the nonlinear activations as a modulating factor in the weight normalisation; its role will be suppressed if the nonlinearity $\phi(x)$ plateaus at large $x$, which corresponds to a small $\phi'(x)$ and thus smaller entries in $\bm{Q}_{2}$. This happens for Tanh and Sigmoid activations, where $\phi_{\text{Tanh}}(x)=\tanh(x)$ and $\phi_{\text{Sigmoid}}(x)=1/(1+e^{-x})$. Thus, we expect that  modularisation behaviour will be suppressed for these activation functions. 

\subsubsection{Effect of data distribution}

The data-dependent nature of $\Delta L$ in Eq.~\eqref{eq:MSE} means that special data distributions may affect the training phenomenology. We briefly explore this here. Consider Eq.~\eqref{eq:MSE} for $\alpha=1$ (Poisson noise) in a linear network without bias:
\eqS
\begin{split}
\Delta L|_{\alpha=1,\text{ linear}}&=\sigma_{n}^{2}\sum_{\mu_{3},\mu_{2}}[\bm{W}_{2}]_{\mu_{3}\mu_{2}}^{2}|[\bm{h}_{2}]_{\mu_{2}}|\\
&=\sigma_{n}^{2}\sum_{\mu_{3},\mu_{2}}[\bm{W}_{2}]_{\mu_{3}\mu_{2}}^{2}\left|\sum_{\mu_{1}}[\bm{W}_{1}]_{\mu_{2}\mu_{1}}[\bm{h}_{1}]_{\mu_{1}}\right|.
\end{split}
\label{eq:MSE3}
\eqE
If the input entries are sparse, for instance if only one entry of $\bm{h}_{1}$ is non-zero in each input sample (one-hot), summing Eq.~\eqref{eq:MSE3} over all different samples reduces the excess loss to a product of $L_{1}$ and $L_{2}$ weight norms, which we call $L_{12}$ regularization:
\eqS
L_{12}\equiv\sum_{\mu_{2}}\left(\sum_{\mu_{3}}[\bm{W}_{2}]_{\mu_{3}\mu_{2}}^{2}\right)\left(\sum_{\mu_{1}}|[\bm{W}_{1}]_{\mu_{2}\mu_{1}}|\right).
\label{eq:L12}
\eqE
Applying $L_{12}$ encourages sparsity for inputs to the hidden neurons $\mu_{2}$, where the input entries are $[\bm{W}_{1}]_{\mu_{2}\mu_{1}}$, by virtue of the $L_{1}$ regularisation on $[\bm{W}_{1}]_{\mu_{2}\mu_{1}}$. The connectivity does not become sparse, because if the neuron has no outgoing weights $[\bm{W}_{2}]_{\mu_{3}\mu_{2}}$ the $L_{1}$ regularisation vanishes. As a result, modularisation should arise even in a linear network in this special case.
\\

\begin{figure}[htb]
\includegraphics[width=8.7cm]{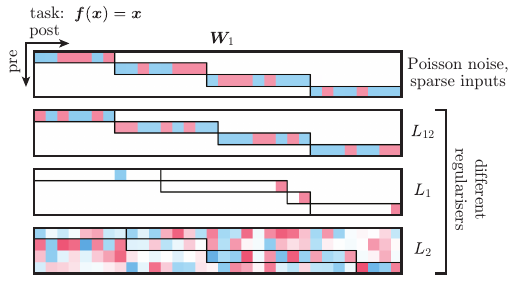}
\caption{\textbf{Mixed ($L_{12}$) regularisation in shallow networks can lead to modular solutions.} We train linear networks with a single hidden layer on the trivial task of learning the identity function $\bm{f}(\bm{x})=\bm{x}$ under different conditions or regularisers. Training with explicit Poisson noise and sparse inputs (top panel), or with the associated $L_{12}$ regulariser without Poisson noise and sparse inputs, Eq.~\eqref{eq:L12}, (second panel) results in modularised connectivity that partitions hidden nodes across inputs. Within each partition the connectivity is dense/all neurons are active. By contrast, $L_{1}$ or $L_{2}$ regularisers (bottom two panels) generate sparse or dense non-modular connectivities, respectively.}
\label{fig:L12}
\end{figure}

We numerically test the effect of Poisson noise injection with sparse training data (only one entry of $\bm{x}$ is non-zero in each example) by training a linear network with a single hidden layer of 32 neurons and 4 inputs and outputs to learn the linear map: $\bm{f}(\bm{x})=\bm{x}$. At convergence, the two weight matrices satisfy $\bm{W}_{2}\bm{W}_{1}=\bm{I}$ where $\bm{I}$ is the identity matrix, so we focus on the structure of $\bm{W}_{1}$ alone. Inspection of $\bm{W}_{1}$ shows a modular connectivity structure (Fig.~\ref{fig:L12}, top panel). The same structure emerges when using the $L_{12}$ regulariser instead of Poisson noise, even if the inputs are not sparse (Fig.~\ref{fig:L12}, second panel). Under the same conditions (non-sparse input), training with $L_{1}$ or $L_{2}$ regularisers does not lead to modularization (Fig.~\ref{fig:L12}, bottom panels), suggesting that $L_{12}$, though it was derived under the special condition of very sparse (one-hot) inputs in single-layer linear networks, may be an interesting regulariser, but its behaviour in deeper networks would require further exploration (Appendix Fig.~\ref{fig:compare3}). Nevertheless, to derive a regularizer under more general conditions, we turn to analyzing the scenario of deeper networks and non-sparse activity. 

\begin{figure*}[htb]
\includegraphics[width=17.8cm]{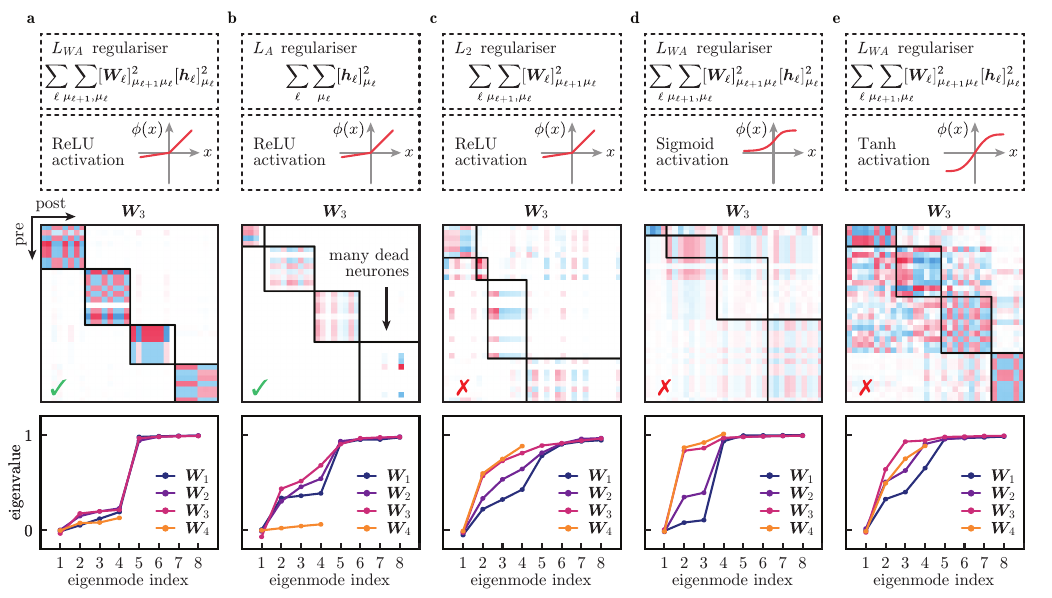}
\caption{\textbf{The weighted-activity  (WA) regulariser $\bm{L_{WA}}$ drives modularisation in deep nonlinear networks.} \textbf{a} The $L_{WA}$ regulariser combines the $L_{2}$ norm and activations, and recapitulates the ANN modularisation behaviour seen with noise injection. \textbf{b} Compared with an activation regulariser, the weighted version avoids dead neurons as no penalty is imposed on neurons that have small input weights. \textbf{c} An $L_{2}$ regularisation does not promote modular connectivity structure. \textbf{d} The error analysis Eq.~\eqref{eq:lossA2NonLin} suggests the neuron activation function acts as a switch on the weight regularisers, and a Sigmoid activation function should have a weaker effect compared to a ReLU. \textbf{e} Similarly, the plateauing behaviour of the Tanh activation does not allow modularisation to emerge either.}
\label{fig:compare}
\end{figure*}

\subsection{Multi-layer networks}

\begin{figure*}[htb]
\includegraphics[width=17.8cm]{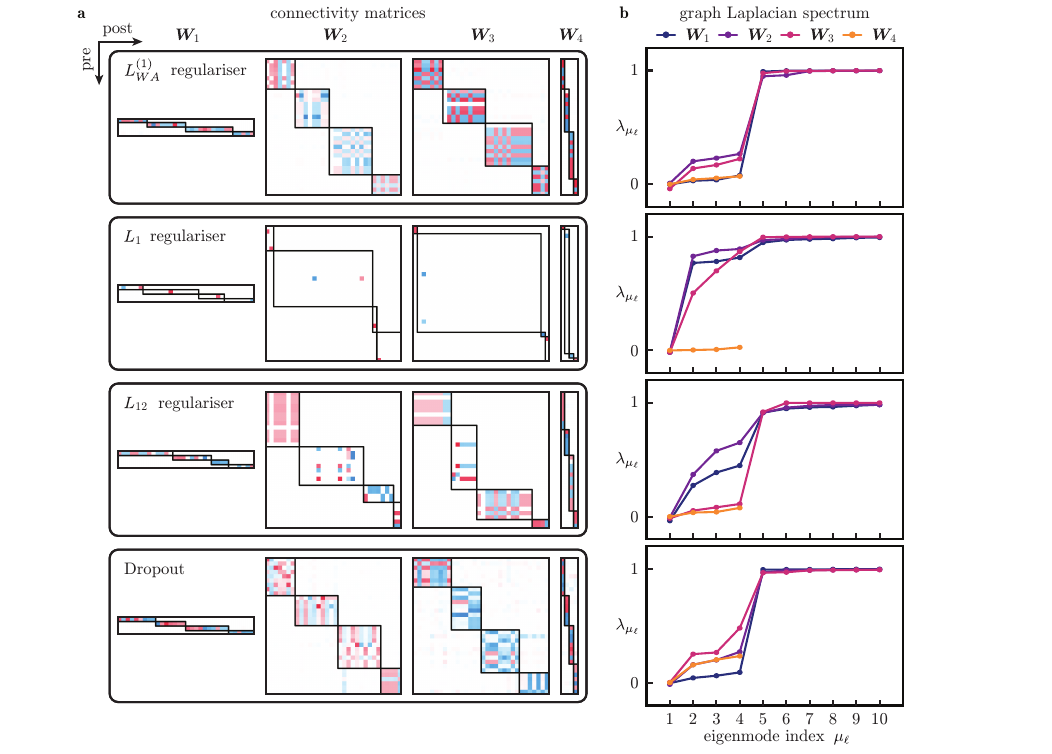}
\caption{\textbf{Effects of other regularisers on network structure.} \textbf{a} Connectivity matrices and \textbf{b} their corresponding graph Laplacian spectra for (top to bottom) $L_{WA}^{(1)}$-regularised, $L_{1}$-regularised, $L_{12}$-regularised networks, and for a network trained with dropout, where the activation of each neuron is set to 0 with probability 0.05 during training. The network connections from $L_{1}$-regularisation are sparse. For $L_{12}$, connectivities are modular but with some dead neurons since it directly suppresses connection weights. Dropout noise, in contrast, has the same variance scaling as multiplicative noise, leading to modular connections.}
\label{fig:compare3}
\end{figure*}

In multi-layer networks, computing the excess loss from noise is complicated because the noise propagates through all the layers. Even if possible to compute, this would result in a complex form of the corresponding deterministic regulariser. Motivated by the observation that modest noise sizes ($\sigma_n=0.1$) are sufficient to drive modularization, and that the effect of noise in a layer will be damped when propagated across layers, we consider an approximate excess loss, based on summing the quadratic errors from each pair of layers. Let 
\eqS
\bm{q}_{\ell+1}\equiv\bm{W}_{\ell}(\bm{h}_{\ell}+\bm{\eta}_{\ell})+\bm{b}_{\ell}
\eqE
denote the input at layer $\ell+1$, before applying the nonlinearity. The quadratic error in $\bm{q}_{\ell+1}$ due to $\bm{\eta}_{\ell}$ is similar to the case of a single hidden layer [cf.~Eq.~\eqref{eq:MSE}]:
\eqS
\begin{split}
\expval{\Delta \bm{q}_{\ell+1}^{T}\Delta \bm{q}_{\ell+1}}_{n}&=\sigma_{n}^{2}\sum_{\mu_{\ell+1},\mu_{\ell}}[\bm{W}_{\ell}]_{\mu_{\ell+1}\mu_{\ell}}^{2}[\bm{h}_{\ell}]_{\mu_{\ell}}^{\alpha}.
\end{split}
\label{eq:regu}
\eqE
We thus define the multilayer network WA loss $L_{WA}^{(\alpha)}$ as
\begin{equation}
L_{WA}^{(\alpha)}\equiv\sum_{\ell}\sum_{\mu_{\ell+1},\mu_{\ell}}[\bm{W}_{\ell}]_{\mu_{\ell+1}\mu_{\ell}}^{2}[\bm{h}_{\ell}]_{\mu_{\ell}}^{\alpha},
\label{eq:L_WA}
\end{equation}
which involves the product of the outgoing
squared weights with the squared presynaptic neural
activations. We add Eq.~\eqref{eq:L_WA} as a regulariser while training a deep nonlinear network on the same task setting as in Fig.~\ref{fig:setup}, and find through the Laplacian eigenspectra and visualisation of the connectivity matrices, that it drives modularity emergence without noise injection for both $\alpha=1$ (Poisson-like noise, Appendix Fig.~\ref{fig:compare3}) and $\alpha=2$ (multiplicative noise) (Fig.~\ref{fig:compare}a). For brevity, we set $\alpha=2$ in the following discussions and write $L_{WA}\equiv L_{WA}^{(2)}$. The WA regularisers promotes modularity in multilayer nonlinear networks even though we did not fully propagate the effective noise errors through the network when deriving the regulariser.
\\

We compare $L_{WA}$ with other regularisers, including the activity loss
\eqS
L_{A}\equiv\sum_{\ell}\sum_{\mu_{\ell}}[\bm{h}_{\ell}]_{\mu_{\ell}}^{2},
\label{eq:L_A}
\eqE
and test the effects of different activation functions. Activity regularization $L_{A}$ can also lead to discovery of a modular solution (Fig.~\ref{fig:compare}b), consistent with earlier findings that penalising neural activity can make ReLU neurons respond to specific task factors \cite{whittingtonDisentanglementBiologicalConstraints2023}. However, it results in `dead neurons' that remain inactive and generates numerical destabilisation of the graph Laplacian (Fig.~\ref{fig:compare}b). With $L_{WA}$, neuron activity regularisation becomes stronger when its outgoing connection weights are larger, and vice versa. This tuning of neuron-level regularisation strength by the readout weight matrix avoids dead neurons since neurons with near-zero weights experience almost no activation penalty and  remain in a position to become active again. 
\\

The normal $L_{2}$ weight regularisation, 
\eqS
L_{2}\equiv\sum_{\ell}\sum_{\mu_{\ell+1},\mu_{\ell}}[\bm{W}_{\ell}]_{\mu_{\ell+1}\mu_{\ell}}^{2},
\eqE
results in no modularisation (Fig.~\ref{fig:compare}c), as predicted in the subsection above. To test our prediction that saturating nonlinearities should thwart modularity, we simulate networks with Tanh and Sigmoid nonlinearities while maintaining the $L_{WA}$ regularisation. Both activations lead to less modularised connectivities than ReLU (Fig.~\ref{fig:compare}d, \ref{fig:compare}e). There is slightly more modular structure with Sigmoid than Tanh neurons, attributable to the fact that tanh reduces to a linear form $\tanh(x)\approx x$ for small $x$, and in the previous subsection we derived that linear responses do not promote modularity.
\\

We additionally show training results with $L_{WA}^{(1)}$, $L_{1}$, and $L_{12}$ regularisers, or with dropout noise, in Appendix Fig.~\ref{fig:compare3}. The $L_{1}$ regulariser promotes high sparsity, leading to unstable graph Laplacian spectra. Applying the $L_{12}$ regulariser from Eq.~\eqref{eq:L12} layer-wise does promote modular connectivity, but also suffers from a moderate degree of dead neurons. We hypothesise that dead neurons appear despite the weighted nature of the $L_{1}$-like suppression in Eq.~\eqref{eq:L12} because only neurons already contributing to functional performance can increase in activity. If the connection weights to a neuron drop to 0 too early in training, the vanishing gradient prevents their re-activation even if that would benefit performance. Finally, dropout noise, where the activation of a neuron is randomly set to 0 in training (probability 0.05 in our simulations), can be treated as an injected noise whose variance scales quadratically with the neuron activation. This is because for an activation $h$, the noise $\eta$ due to dropout is $-h$ with probability $p=0.05$ and $0$ otherwise, leading to $\expval{\eta^{2}}=(-h)^{2}\cdot p+0\cdot(1-p)=h^{2}p$, so it also promotes modular connectivity, similar to multiplicative noise.

\subsection{Simultaneous discovery of modular basis and modular solution}

The WA regularisation discussed above promotes the emergence of fully modular connectivity in tasks that involve modular functions of the inputs, where the inputs are already disentangled -- they are already the correct basis in which the tasks are modular. But if the task requires a step that mixes across the modular streams, can the regularizer $L_{WA}$ still find an appropriate modular solution? Relative to the required mixed inputs, the unmixed network inputs are effectively entangled; thus the network has to simultaneously find the correct mixing/demixing of the inputs to find the basis in which there is a modular solution while finding the modular solution. 
\\

\begin{figure}[tb]
\includegraphics[width=8.7cm]{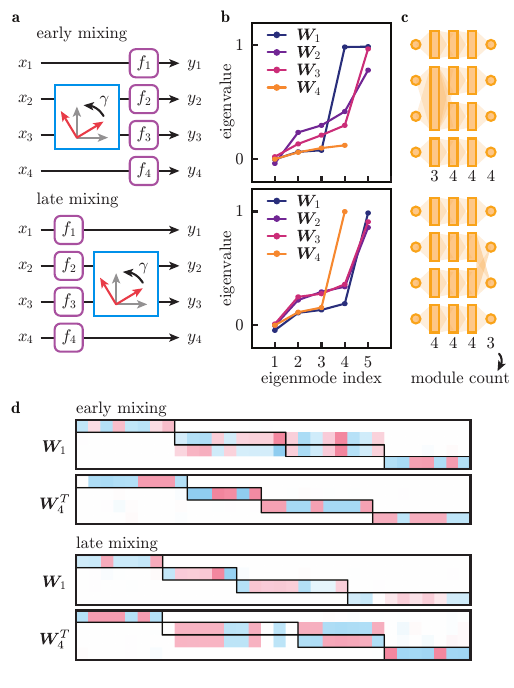}
\caption{\textbf{$\bm{L_{WA}}$ drives ANNs to discover demixed independent task factors within the fewest possible layers.} \textbf{a} We compare two learning objectives under the WA regularisation Eq.~\eqref{eq:L_WA}: early mixing of inputs according to Eq.~\eqref{eq:mixEarly} (top) and late mixing of outputs from nonlinear functions according to Eq.~\eqref{eq:mixLate} (bottom). The mixing function is a linear operator represented by the blue square, and nonlinear functions are in purple. \textbf{b} Examining the Laplacian matrix of network connectivities in reveal that in the early mixing case (top), the first layer has a 3-module structure and later layers have 4 modules, indicating the first layer decomposes the two middle inputs and the later layers then compute the nonlinear functions $f_{i}(x)$. In contrast, for late mixing (bottom), the first three layers have 4 modules that compute the $f_{i}(x)$'s on their own, and the outputs from the middle channels are mixed only at the final layer. \textbf{c} Schematic of the distinct modularisation behaviour, number of modules counted from the Laplacian spectra are stated at the bottom. \textbf{d} Connectivity matrices of the first and last layers of these networks, showing channel cross-talks in the first layer for early mixing (top) and the last layer for late mixing (bottom). Other intermediate layers all have a 4-cluster blocky structure. 
}
\label{fig:demix}
\end{figure}

\begin{figure*}[tb]
\includegraphics[width=17.8cm]{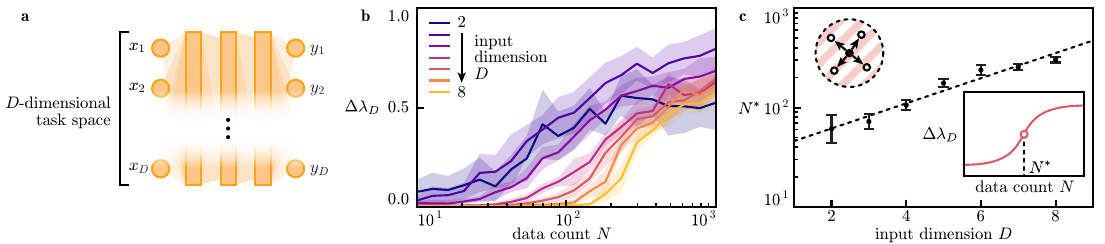}
\caption{\textbf{Sample complexity of  modularity emergence through WA regularization.} \textbf{a} To study the data complexity of network modularisation under $L_{WA}$-regularisation, we train networks to learn $D$-dimensional modular functions. \textbf{b} The eigengap $\Delta\lambda_{D}$ corresponding to a $D$-module connectivity shows a sigmoidal growth trend with $N$ (number of unique data samples per epoch) on a logarithmic scale. 
\textbf{c} For each curve, we identify as $N^{*}$ the critical data count required for a network to adopt a modular structure (right inset), and an exponential scaling of $N^{*}$ with input dimension $D$ is revealed. This scaling is consistent with a picture where modular connectivity emerges when training data points come within some critical distance to one another, such that the local function gradient can be estimated and a modular task structure discovered by the network (left schematic). Dashed line is an exponential scaling reference.
}
\label{fig:dataN}
\end{figure*}

Therefore we consider tasks in which as before the target outputs involve modular functions $\bm{f}=[f_{1}(x_{1}),f_{2}(x_{2}),f_{3}(x_{3}),f_{4}(x_{4})]^{T}$, but also involve a mixing step: 
\begin{subequations}
\begin{align}
\bm{y}_{\text{early}}\equiv\bm{f}(\bm{R}\bm{x}),\label{eq:mixEarly}\\
\bm{y}_{\text{late}}\equiv\bm{R}\bm{f}(\bm{x}).\label{eq:mixLate}
\end{align}
\end{subequations}
where the matrix $\bm{R}$ mixes two of the independent factors together. The `early' mixing target requires inputs to be mixed before application of the independent functions (Fig.~\ref{fig:demix}a, top); the `late' mixing target requires computation of the modular functions followed by mixing their outputs  (Fig.~\ref{fig:demix}a, bottom). \\

In simulations we picked a mixing matrix that rotates factors 2,3 by angle $\gamma=\pi/4$: 
\eqS
\bm{R}\equiv\begin{pmatrix}
1&0&0&0\\
0&\cos\gamma&\sin\gamma&0\\
0&-\sin\gamma&\cos\gamma&0\\
0&0&0&1
\end{pmatrix},
\eqE
and trained identical fully connected ANNs with three hidden layers and the WA regulariser $L_{WA}$ on these two tasks. Strikingly, all layers of the trained networks became modular in both cases, with most layers exhibit a 4-module structure, suggesting they have discovered that 4 independent nonlinear factors (Fig.~\ref{fig:demix}b) comprise the task. For the early mixing task, the first hidden layer $\bm{W}_{1}$ acquires a 3-module connectivity (3 near-0 eigenvalues and the 4th eigenvalue closer to 1, Fig.~\ref{fig:demix}b-d, top). Subsequent layers then reflect the fully modular downstream aspect of the early mixing task. In other words, the network for early mixing task mixes as required the independent factors, doing so at the earliest possible point and only at that point, and then all subsequent layers develop a fully modular unmixed structure. \\

In contrast, for the late mixing task, the network creates a fully modular propagation and function computation on the four independent inputs through its layers but the last, and performs the required mixing of the functions in only the last layer (Fig.~\ref{fig:demix}b-d, bottom). If we consider the regulariser as mimicking the effects noise injection, then the network behaviour we found here indicates that the mixing of variables incurs excess noise error so that the network works to maximally disentangle and keep independent task factors across its layers, mixing only when necessary and doing so within the fewest layers. Conversely, one could use this observation to infer, in general tasks, the degree of nonlinear processing in the training data by examining how an ANN with $L_{WA}$ regularisation de-mixes task factors. 
\\

\section{Sample complexity for discovery of modular structure}

We have established that the WA regulariser drives the discovery of rare solutions with modular network connectivity after training while networks without WA regularization do not. However, we do not yet know the sample complexity of this discovery process. We next ask how rapidly modular structure is discovered and acquired, by investigating the sample complexity of modularity emergence with the WA regularizer as a function of $D$, the number of input dimensions (so far we used $D=4$), Fig.~\ref{fig:dataN}a. \\

We train networks with varying $D$ (the hidden layers scale linearly in size with $D$) and as before quantify network modularity by the $D$-th eigengap $\Delta\lambda_{D}$. We plot the value of this gap as a function of the number of training datapoints $N$ (Fig.~\ref{fig:dataN}b). Modularity emerges later for larger $D$ (Fig.~\ref{fig:dataN}b); we fit $\Delta\lambda_{D}$ to the sigmoidal function $\Delta\lambda_{D}=\Delta\lambda_{D}^{\text{max}}/\{1+\exp[-v(N-N^{*})]\}$ with steepness $v$ and transition point $N\sim N^{*}$ with a rate $v$ (Fig.~\ref{fig:dataN}c, right inset). Plotting the resulting transition point $N^{*}$ against $D$  reveals an exponential scaling of number of training data points $N^*$ for modularity emergence as a function of number of modular task dimensions (Fig.~\ref{fig:dataN}c). 
\\

To understand the origins of exponential scaling in modularity emergence, we consider that the transition from a non-modular network structure to a modular one in a flexible learner capable of learning tasks that are non-modular or modular of any dimension $\leq D$ requires that it infer from the training data whether and which  function gradients are zero, $\partial_{j}f_{i}\equiv\partial f_{i}/\partial x_{j}=0$ for $i\neq j$. Around a reference training point $x_{i}^{1}$ (with target $y_i^1$), another training point $x_{i}^{a}$ (within a hypercube of side length $\Delta x^*$ of the reference, with target $y_i^a$) provides information on the first-order gradients via $y_{i}^{a}-y_{i}^{1}\approx\sum_{j}[\partial_{j}f_{i}(x^{a})-\partial_{j}f_{i}(x^{1})](x_{j}^{a}-x_{j}^{1})\approx \sum_{j}\partial_{j}f_{i}(x^{1})\Delta x_{j}^{a}$, assuming that the terms $\Delta x_{j}^{a}$ (and therefore $\Delta x^*$) are small enough that the higher-order terms such as $\partial_{i}^{2}f_{i}$ vanish. Within this first-order approximation, local gradients can be estimated by solving the simultaneous equations $\Delta y_{i}^{a}=\sum_{j}\Delta x^{a}_{j}\partial_{j}f_{i}(x^{1})$, which requires (pseudo)inversion of the data matrix $\Delta x^{a}_{j}$. This matrix attains rank $D$ only if there are at least $D$ training points in the hypercube; in addition, higher-order terms act as a source of noise under the linear approximation, so it would require a scaled factor of $\beta >1$ more data samples, where $\beta$ increases with the effective noise and inversely with the desired squared precision of gradient estimation. Thus, we need $\sim \beta D$ data points within each hypercube of length $\Delta x^{*}$. Suppose the full domain is a hypercube of length $L$, the expected number of points within the small hypercube is $N^{*}\cdot(\Delta x^{*}/L)^{D}$. Equating the two, we get $N^*\sim \beta D\cdot(L/\Delta x^*)^D$. 
\\

The sample complexity of modularisation by a fully flexible learner (capable of learning any function) is therefore exponential in $D$ to leading order [$\mathcal{O}(Dm^{D})$ where $m$ denotes the number of points needed for one dimension, $D=1$]: this is the curse of dimensionality, which arises from the fundamental requirement of gradient sensing, in which a minimum number of samples is required per dimension to determine whether there is a non-zero or zero gradient in that direction \cite{boopathyBREAKINGNEURALNETWORK2025}. Deep network learning with the WA modularity regularizer remains flexible; though it promotes discovery of the modular solution when one exists, from the vast set of functions that satisfy the task, it nevertheless remains capable of learning non-modular solutions. The argument above and the results from Fig.~\ref{fig:dataN}c show that WA regularization is distinct from making a learner a priori aware of the specific modular structure of the task. Given that a fully modular task is in principle equivalent to $D$ copies of single-input, single-output function-learning tasks, a learner fully aware of its specific modular structure could optimally learn the task with a sample complexity as small as $\mathcal{O}(m)$. In sum, the WA regularizer enables the discovery of task modularity and finds the correct modular solution, but does not solve the curse of dimensionality in doing so. \\

Once modularity has been established, however, other functional benefits ensue, as we explore next.

\begin{figure}[tb]
\includegraphics[width=8.7cm]{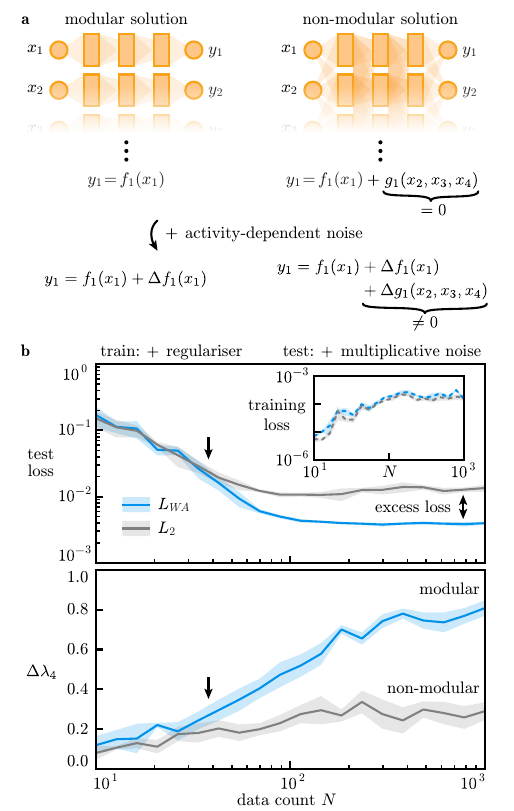}
\caption{\textbf{Noise-robustness from modularity.} \textbf{a} An output node from a modular network depends only on the relevant input, but that from a non-modular network can have dependencies on other inputs that cancel out, represented by an additional term $\epsilon_{2}(x_{2},x_{3},x_{4})$ that is identically 0. However, in the presence of noise, this extra term can lead to non-zero fluctuations. \textbf{b} We train networks with WA or L2 regularisers, with a varying number of training points $N$, and test the performance with explicit multiplicative noise in activation layers of strength $\sigma_{n}=0.1$. WA-regularised networks outperform L2-regularised ones with sufficient $N$ (top) and this occurs in tandem with the emergence of connectivity modularity, indicated by an increase in the 4th eigenvalue gap of the first connectivity weight matrix (bottom). 
}
\label{fig:robustnessMain}
\end{figure}

\section{Functional benefits of network modularity}

\subsection{Noise robustness}

Since the WA regulariser is derived from effective errors in the presence of activation noise, we expect $L_{WA}$-regularised networks to perform better under noise injection even though they are trained without noise, compared to networks trained without the WA regulariser. Mechanistically, each output node of a trained modular network is a function of a single variable, for instance $y_{1}=f_{1}(x_{1})$ for the first output node. In contrast, a non-modular network computes functions of irrelevant inputs but attains the same performance by learning that the value of the parts of the function that depend on irrelevant variables in the training data are zero: $y_{1}=f_{1}(x_{1})+g_{1}(x_{2},x_{3},x_{4})$ with $g_{1}(x_{2},x_{3},x_{4})=0$. In the presence of noise during testing, the cancellation of the irrelevant inputs in $g_1$ becomes non-exact and the non-modular network output acquires the extra term $\Delta g_{1}(x_{2},x_{3},x_{4})>0$, which acts as an additional source of error (Fig.~\ref{fig:robustnessMain}a). We characterise and compare the noise robustness of $L_{WA}$-regularised and $L_{2}$-regularised networks when training over different amounts of data. The regularisers are applied during training, and activation noise of strength $\sigma_{n}=0.1$ and scaling exponent $\alpha=2$ (multiplicative noise) is injected to the network during testing. Indeed, as the number of training data points $N$ increases and the $L_{WA}$-regularised network becomes more modular (Fig.~\ref{fig:robustnessMain}b, bottom panel), in tandem the test loss of the $L_{WA}$ network begins to drop below that of the $L_{2}$-regularised network (Fig.~\ref{fig:robustnessMain}b, top panel).

\subsection{Modular architecture drives disentangled processing and compositional generalisation}

\begin{figure*}[htb]
\includegraphics[width=17.8cm]{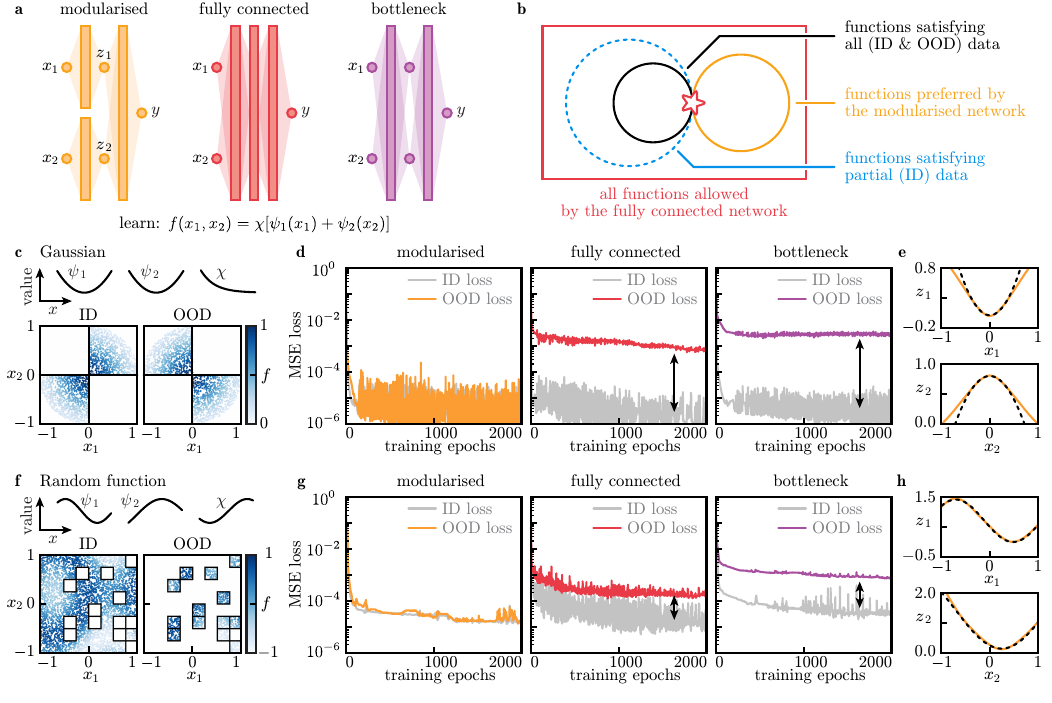}
\caption{\textbf{Modularity-favoring architectural bias for strong out-of-distribution compositional generalisation.}
\textbf{a} Schematics of the architectures tested. Each circle represents a single neuron unit, and the rectangles represent a group of neurons. \textbf{b} The data distribution and network architecture place constraints on the function achievable by the ANN, and generalisation occurs when the constraint spaces only intersect each other on low-dimensional manifolds. \textbf{c} The Gaussian task requires an ANN to learn a 2-dimensional Gaussian function, and we split the full domain into a training (in-distribution, ID) part and a testing (out-of-distribution, OOD) part, each comprising two quadrants within a unit circle. \textbf{d} Testing different network architectures reveals that the modularised network (left) has learned the Gaussian function across the whole domain even with partial data, since the ID and OOD loss completely overlap. In comparison, the full (middle) and bottleneck (right) networks still exhibit sizable generalisation gaps. \textbf{e} The compositional generalisation capability of the modularised network stems from individual modules learning quadratic functions up to additive and multiplicative gauge factors. Orange lines are the outputs of the modules, dashed lines are quadratic functions. \textbf{f} We test the generalisation bias in a more generic task setting where the single-variable functions are randomly generated. The ID and OOD regions are unions of randomly selected, non-overlapping patches. \textbf{g} Plotting the ID and OOD losses show that the same generalisation phenomenon is recovered as in the Gaussian task. \textbf{h} Module outputs agree with the component functions $\psi_{1}$ (top) and $\psi_{2}$ (bottom), upto scaling and an additive constant.}
\label{fig:Gaussian}
\end{figure*}

\begin{figure}[b]
\includegraphics[width=8.6cm]{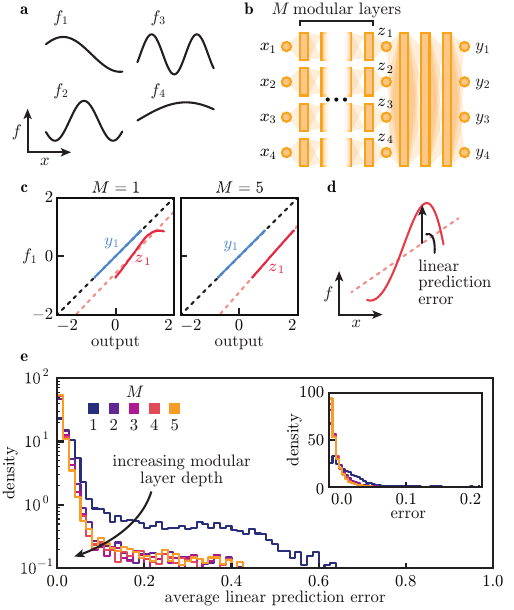}
\caption{\textbf{Modularised blocks are predisposed to encode non-linear transformations.}
\textbf{a} We revisit the task of learning 4 independent single-variable functions. \textbf{b} The modularised network contains $M$ layer-deep modularised blocks with intermediate outputs $z_{1}$, $\dots$, $z_{4}$, and a fully connected block of fixed depth $H=3$. \textbf{c} For small $M$ the modular block outputs non-linearly deviate from the desired outputs (deviation of solid red curve from dashed red curve). For larger $M$, the modular blocks output a response that is proportional to the final output $y_{1}$ meaning they have learned the modular functions up to linear scaling. \textbf{d} Quantifying the deviation between the nonlinear modular target functions and the output of the modular blocks by fitting a straight line in a plot of target value and output. The deviation from this linear predictor reflects the residual error. \textbf{e} Histograms of the errors from {\bf d} while parameterically varying $M$, for different functions and random seeds, showing smaller errors in the modular outputs as $M$ increases. Inset: same plot in linear scale.}
\label{fig:Mod4Fun}
\end{figure}

We next explore the effects of modularity as a pre-existing architectural bias --- which we may view as established through evolution or through prior learning on related tasks with a WA-type regularizer --- and show that modular information processing channels enable learning with far out-of-distribution generalisation through compositionality. We consider learning of 2-input, 1-output functions of the form
\begin{equation}
f(x_{1},x_{2})=\chi[\psi_{1}(x_{1})+\psi_{2}(x_{2})],
\label{eq:fx1x2}
\end{equation}
where $\chi$, $\psi_{1}$, and $\psi_{2}$ are single-variable functions. As universal function approximators \cite{Debao1993,Zhou2020c,Pinkus1999,Augustine2024,Leshno1993}, any network with 2 inputs, 1 output, and reasonable size should be able to learn Eq.~\eqref{eq:fx1x2}. To test qualitative effects of network architectures, we compare three different feedforward ANN architectures: modularised, fully-connected (FC), and bottleneck networks (Fig.~\ref{fig:Gaussian}a). Only weight decay is used as a regularisation method here. The inputs in the modularised network project disjointly to the first hidden layer, then are combined in the downstream layer. The FC network has the same number of units and hidden layers but is fully connected (no architectural constraint). The bottleneck network resembles the modularised network, except that the first hidden layer is FC instead of disjoint.
\\

We hypothesise that the modularised network is biased toward learning the true single-variable component functions $\chi$, $\psi_{1}$, and $\psi_{2}$, while the FC and bottleneck networks memorise the output value given an input. If so, then when the input to the trained networks is drawn from an out-of-distribution (OOD) domain with no in-distribution (ID) support, the modularised network should be able to generalise but the other networks would not. Conceptually, consider the (very high-dimensional) space of all functions (Fig.~\ref{fig:Gaussian}b). The modularised network is equivalent to the FC network with certain weights constrained to 0, thus the set of functions allowed by the modularised network is a subset of those representable by the FC network (Fig.~\ref{fig:Gaussian}b, red and orange regions). Now consider the data distribution. Each dataset places constraints on candidate functions. The full dataset (ID and OOD) specifies a smaller subset of functions than the ID dataset (Fig.~\ref{fig:Gaussian}b, black and blue, respectively). If trainable on a specific dataset, a network with a specific architecture will converge to a function in the intersection of the architectural subset and data constraint subset. Ideally, for the modularised network these sets would intersect at a low-dimensional manifold representing the ground truth function (Fig.~\ref{fig:Gaussian}b, star), allowing generalisation to occur before the full dataset is presented to the network. \\

To test this, we train networks to learn a 2-dimensional Gaussian function:
\begin{equation}
f(x_{1},x_{2})=\exp\left[-\frac{1}{2\sigma^{2}}(x_{1}^{2}+x_{2}^{2})\right],
\end{equation}
with $\sigma=1/2$. The networks are trained on the ID dataset comprising all points satisfying $x_{1}x_{2}>0$ within the unit circle $x_{1}^{2}+x_{2}^{2}<1$, while the OOD is the complementary set where $x_{1}x_{2}<0$, also within the circle (Fig.~\ref{fig:Gaussian}c). When we compute the mean squared error loss for ID and OOD sets as the networks are trained, we observe that the OOD loss nearly matches the ID loss for the modularised network (Fig.~\ref{fig:Gaussian}d, left), suggesting it has learned the component functions, while the FC and bottleneck architectures  exhibit a large generalization gap (Fig.~\ref{fig:Gaussian}d, middle and right). To further verify that the modularised network has learned the component functions, we plot the intermediate outputs $z_{1}$, $z_{2}$ from the modular blocks (Fig.~\ref{fig:Gaussian}a, \ref{fig:Gaussian}e). For a Gaussian function, these should be quadratic functions of the inputs $x_{1}$ and $x_{2}$, and indeed the modular block outputs are quadratic functions (Fig.~\ref{fig:Gaussian}e), up to a multiplicative and additive coefficient, which represent gauge freedoms from the weights and biases of the downstream layer. In addition, we test the effect of different ID/OOD split by parameterising these sets as fan-shaped regions, and find that the modularised network learns the true function with much less ID data than the other networks (Appendix \ref{A:alter}).\\

To test the generalizability of this result, that a modular network learns the modular structure of functions, we train these networks on generic modular function composition tasks. Each task selects component functions $\psi_{1}$, $\psi_{2}$, and $\chi$ from a set of sinusoidal waves with randomly sampled frequencies and phases. ID and OOD sets here are non-overlapping random patches (Fig. ~\ref{fig:Gaussian}f). Consistent with the Gaussian result, learning this task with the modularized network results in excellent OOD generalization, while the other architectures do comparably on ID data but much worse on OOD data (Fig.~\ref{fig:Gaussian}g). When we examine what functions are learned in which layers, we find that the modules in the modular layer learn the correct nonlinear functions up to scaling (Fig.~\ref{fig:Gaussian}h).\\

Having demonstrated the inductive bias of modularised networks to learn component functions, we further ask what causes the network to selectively localise modular functions learning in the modular layers, as in the results of Fig.~\ref{fig:Gaussian}, rather than smearing the modular function computation across all layers. If a network has modularised blocks before the FC layers, it could in principle perform most processing in the FC layers and implement identity functions in the modular layers. We examine this question in more detail by returning to the modular function-learning task of Eq.~\ref{eq:modulartask} (Fig.~\ref{fig:Mod4Fun}a), using a 4-function task on 4 inputs and 4 modules (Fig.~\ref{fig:Mod4Fun}b), with 3 FC layers after the modular blocks. We study how the network partitions the computation of the individual functions across layers as we vary the depth $M$ of the modular blocks . The total number of neurons in each modular layer is the same as the number in each FC layer. 
\\

After verifying that the network has learned the target functions (by plotting the outputs $y_{i}$ against the targets $f_i$, Fig.~\ref{fig:Mod4Fun}c, blue), we compare the modular block outputs $z_{i}$ to the target functions $f_{i}$ Fig.~\ref{fig:Mod4Fun}c, red). We find that the FC layers perform some of the non-linear transformations when the modular blocks are shallow (e.g. $M=1$, Fig.~\ref{fig:Mod4Fun}c, left, compare nonlinear red solid line with linear dashed), but when the modular blocks are deeper (e.g. $M=5$) they  learn the modular structure of the task and each module fully represents a nonlinear target function (upto linear scaling, Fig.~\ref{fig:Mod4Fun}c, right). We quantify the degree to which the subsequent FC layers implement the non-linear transformations by the root-mean-squared deviation of the $f_{i}-z_{i}$ curve (Fig.~\ref{fig:Mod4Fun}c-d, red curve) from the best-fit line (Fig.~\ref{fig:Mod4Fun}c-d, red dashed curve). 
Computing this deviation from linear for many different functions and random seeds, we observe less deviations as we parametrically increase $M$ (Fig.~\ref{fig:Mod4Fun}e, Appendix Fig.~\ref{fig:ModSweep}). Interestingly, there seems to be an abrupt transition: once the modular blocks are sufficiently expressive (i.e. with enough depth, which in this case is $M=2$), nonlinear modular function learning localizes to them, even if the FC layers are deeper (overlapping curves for $M>2$ in Fig.~\ref{fig:Mod4Fun}e, Appendix Figs.~\ref{fig:ModSweep} and \ref{fig:DataHMSweep}). Furthermore, though the FC layers learn part of the non-linear modular target functions when the modular layers are shallow ($M<2$) and the proportion of networks with the FC layers learning part of the nonlinear modular targets increases as the FC layers increase in depth, this stops when $M\geq 2$: then, varying the FC layer depth does not change the proportion of networks where the FC layers compute part of the non-linear modular targets (Appendix Fig.~\ref{fig:DataHMSweep}). The analytical basis of this finding can be an interesting direction for future exploration. \\

In sum, we have empirically found another inductive bias of the modular architectural constraint: in the presence of both modular and fully connected blocks, independent non-linear operations are preferentially encoded in the modular blocks when those are sufficiently expressive (above a critical depth that depends on the targets but not on subsequent FC layer depth) even though in principle they could have been localized to the FC layers or spread across both. 

\section{Discussion}

While noise injection has been viewed as a mechanism to perform gradient estimation \cite{williamsSimpleStatisticalGradientfollowing, seungLearningSpikingNeural2003, fieteGradientLearningSpiking2006} or to prevent overfitting \cite{srivastavaDropoutSimpleWay}, our work demonstrates that its influence is deeper, shaping the emergent structural properties of ANNs.  Specifically, we showed that non-additive, activation-dependent noise (e.g., Poisson or multiplicative) leads to the spontaneous emergence of modular structure to solve modular tasks in nonlinear networks. Characteristics of the nonlinear neuron activation functions are crucial for this process too. This result parallels longstanding insights in neuroscience, where neural noise is not simply tolerated but is hypothesised to support robustness, fault tolerance, and even computation itself \cite{faisalNoiseNervousSystem2008}. Our findings suggest that similarly, in artificial systems, noise can serve as an inductive bias toward modularity, echoing prior computational work on noise-driven structure emergence \cite{lorenzEmergenceModularityBiological2011, mccourtNoisyDynamicalSystems2023}.
\\

By analytically examining the excess loss induced by noise in a single-layer network, we derived the WA regulariser as a deterministic analogue mechanism to  noise-driven modularity. This regulariser penalises the product of weights and activations, capturing the modularising effects of Poisson-like or multiplicative activation noise in a deterministic form. Unlike a standard activation regulariser, WA regularisation does not suppress neurons indiscriminately---rather, it naturally avoids `dead neurons' by only penalising active, highly connected units. 
\\

The WA regulariser drives the discovery of exponentially rare modular solutions when the task is modular, but allows the network to remain a fully flexible learner of non-modular tasks as well. As a result, the discovery and emergence of a modular solution requires a number of training data points that scales exponentially with the dimensionality of the problem. We next showed that starting with modular connectivity promotes disentangled downstream information processing, a hallmark of compositional learning. We demonstrated that networks with initial modular layers provide a bias for modular processing, enabling  networks to extrapolate correctly far out of distribution from partial data, outperforming unconstrained architectures by learning shared substructures that generalise beyond the training distribution. This mirrors phenomena in symbolic reasoning and generative modeling, where compositional generalisation hinges not just on disentanglement but on appropriate architectural priors, showing strong out-of-distribution generalization and sub-exponential data scaling \cite{liangCompositionalGeneralizationRequires2025}, when a modularized network architecture is trained on a modular generative task. This agreement of results suggests that constrained architectures that force inputs to be independently processed or ``rendered'' could potentially be a generic strategy for enhancing the generalisation capabilities of neural networks. Further, quantifying the data scaling behaviour of our tasks as a function of task dimension would provide insight into how the model learns. Specifically, we hypothesise based on results in \cite{liangCompositionalGeneralizationRequires2025} that as the modular function learning task dimension increases, the amount of data required by modularized architectures will scale sub-exponentially with the number of dimensions. Thus, a modular architecture might also provide the appropriate inductive bias to solve the curse of dimensionality for modular problems; because our results did not depend on the depth of the downstream FC layers, it is possible that this benefit could be derived without a large loss in expressivity.\\

These findings --- on the advantages of modularity once present but slow acquisition --- may offer insights into the developmental path of the brain's neural networks. Modularised circuits for sensory, motor, and cognitive processing are widely observed in the brain \cite{luRevealingDetailVisual2018, hulseMechanismsUnderlyingNeural2020, lyuBuildingAllocentricTravelling2022, haftingMicrostructureSpatialMap2005,burakAccuratePathIntegration2009}. Many of these circuits are formed in utero or in the early post-natal period, and then their structures remain largely fixed; these modularized architectures are then re-used across the lifetime of the individual as they face diverse tasks. Given our observation on the exponential sample-complexity of discovery of modular solutions by a flexible learner, but the strong advantages of learning modular tasks by modular learners, it appears that biology uses evolutionary time-scale information to discover the modular structure of problems, and then packs and unpacks that evolutionary knowledge into genetically- and spontaneous activity- guided unfolding of developmental programmes that create modular structures over development, to be used in new learning by the adult individual. 
\\

We have seen that modular networks effectively bias learning toward structured factorized functions when the task is modular. More broadly, the Kolmogorov-Arnold representation theorem proves that any continuous real-valued multivariate functions can be factorized into a superposition of continuous single-variable functions 
\cite{kolmogorovRepresentationContinuousFunctions1963, schmidt-hieberKolmogorovArnoldRepresentation2021}:
\eqS
f(x_{1},x_{2},\dots,x_{n})=\sum_{q=1}^{2n+1}\chi_{q}\left[\sum_{p=1}^{n}\psi_{pq}(x_{p})\right].
\eqE
In related work, the  Kolmogorov–Arnold representation theorem prompted the search for alternative neural network architectures, known as KANs, in which edges learn univariate nonlinear functions and nodes perform simple sums \cite{duganOccamNetFastNeural2023, liuKANKolmogorovArnoldNetworks2025,Cranmer2020}. These models can perform interpretable symbolic regression with appropriate regularization and pruning. Our work may be viewed as dual to the KAN models, showing that conventional neural networks can learn solutions that match the modular structure of an actual modular problem without explicit pruning, and that additionally they exhibit far greater out-of-distribution generalisability than fully connected networks when there is limited training data. The Kolmogorov-Arnold theorem further suggests that it may be possible to use modular architectures as we have done here to learn intepretable, generalizable approximations to even apparently non-modular functions. These results lead to important questions for further exploration. What is the relationship between architectural modularity and sample complexity in learning an appropriately modular task? How can we characterise the functional classes allowed by a specific architecture? Future analytical study of how and when the function decomposition transition occurs in our networks could guide the deployment of modular architectures for more performant, generalizable, and explainable networks.\\

In sum, our work shows how Poisson or multiplicative noise, or equivalent deterministic WA regularisers, drive neural networks to discover the vanishingly small space of modular solutions out of the space of all solutions to modular problems, complementing and adding to a recent finding in Boolean networks \cite{mccourtNoisyDynamicalSystems2023}. We also offer an explanation for the prevalence of modularity in biological brains and a path toward embedding such inductive biases in artificial systems to support more robust, interpretable, and generalisable learning.
\\

\textbf{Code availability.} Simulation scripts can be found on the GitHub repository \cite{dq219HttpsGithubcomDq2192025b}.
\\

\textbf{Acknowledgments.} D.Q.~is supported by Transition Bio Ltd.~and the K.~Lisa Yang ICoN Center. I.R.F.~is supported in part by the Simons Foundation (SCGB program 1181110), the ONR (award N00014-19-1-2584), and the NSF (CISE award  IIS-2151077 under the Robust Intelligence program). The authors thank Dr.~Liu Ziyin for helpful comments.
\\

\balancecolsandclearpage

\appendix


\section{ID/OOD split in the Gaussian task}
\label{A:alter}

\begin{figure*}[b!]
\includegraphics[width=17.8cm]{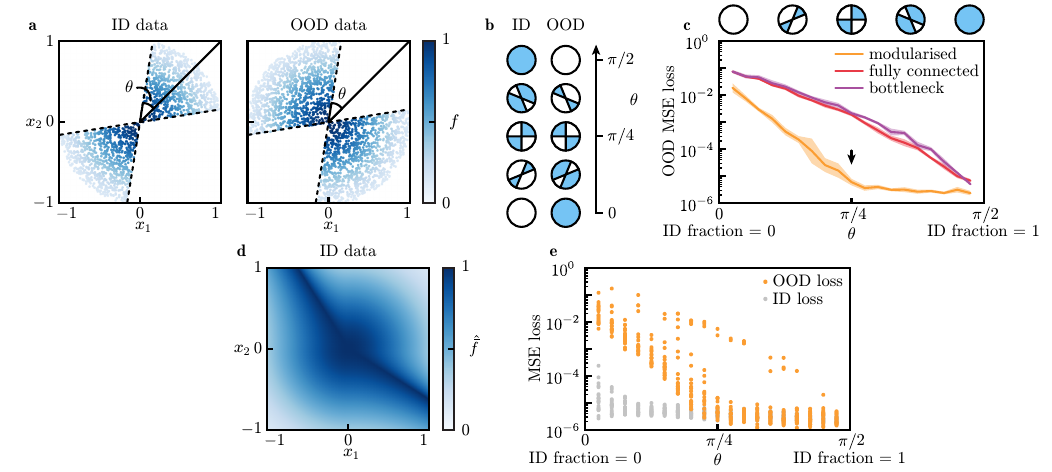}
\caption{\textbf{Data efficiency in the Gaussian function-learning task.} \textbf{a} We construct ID and OOD datasets as fan-shaped regions parameterised by the angle $\theta$. \textbf{b} As $\theta$ varies from 0 to $\pi/2$, input data domain grows to eventually cover the full unit disk, and the OOD region shrinks. \textbf{c} Plotting the terminal OOD loss for the three different network architectures shows that only the modularised network is able to generalise far into the OOD region, at $\theta=\pi/4$ (arrow). \textbf{d} The alternate solution $\hat{\bar{f}}$ given by Eqs.~\eqref{eq:altersol} and Eq.~\eqref{eq:alterOverall}, where it is a usual 2D Gaussian in upper right and lower left (ID) quadrants, but has a ridge in the other (OOD) quadrants. \textbf{e} Scatter plot of losses for all networks sampled in Fig.~\ref{fig:Gaussian}f. Around 2 and 3 networks converge to the alternate solution out of 20. In contrast, all sampled networks for the other architectures follow the group averages in Fig.~\ref{fig:Gaussian}f.}
\label{fig:theta}
\end{figure*}


\begin{figure*}[b!]
\includegraphics[width=17.8cm]{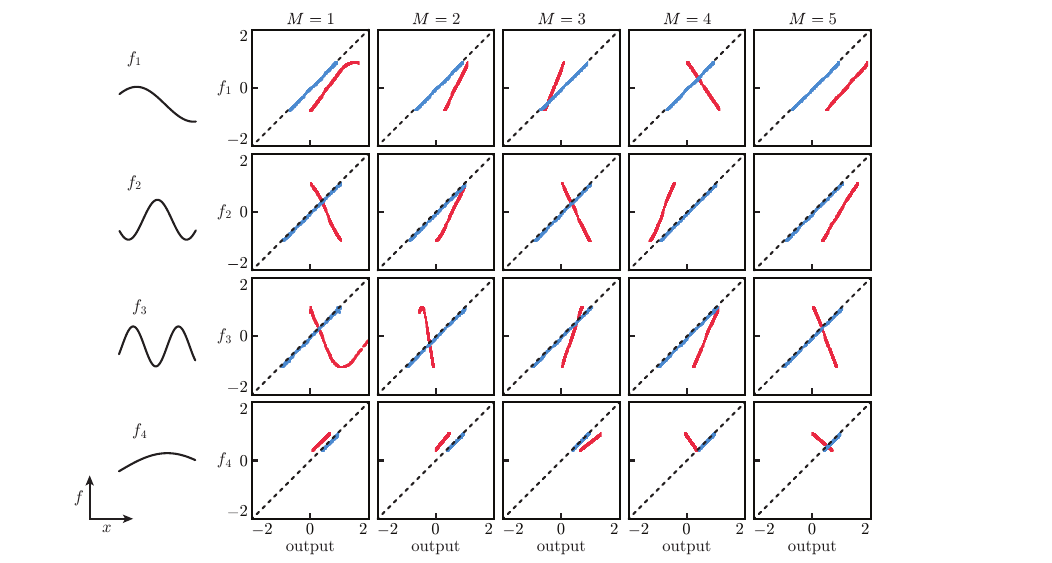}
\caption{\textbf{Several examples of intermediate modular outputs.} The scatter plots are from the simulation of Fig.~\ref{fig:Mod4Fun}, with 3 fully connected hidden layers and varying modular block depth $M$. Even with $M=2$, most of the modular nonlinear functions are fully learned in the modular layers (straight red lines for most of the functions $f_i$).}
\label{fig:ModSweep}
\end{figure*}

\begin{figure*}[b!]
\includegraphics[width=17.8cm]{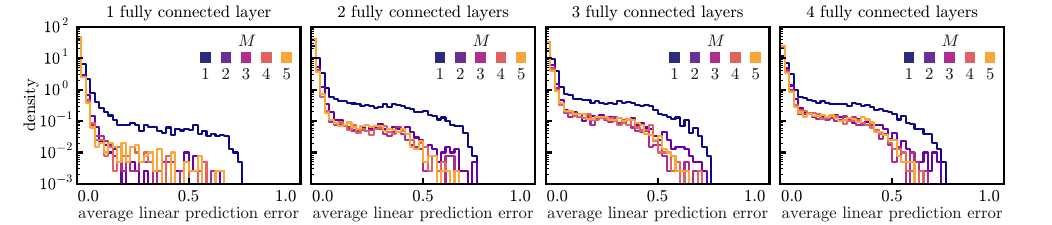}
\caption{\textbf{Distributed expressivity to modularity transition in pre-modularised networks.} We repeat the simulations in Fig.~\ref{fig:Mod4Fun} with a varying number of fully connected layers. Fig.~\ref{fig:Mod4Fun}e is identical to the panel with 3 fully connected layers. We also normalise the histograms by the fraction of simulation runs that actually converged (out of 5000), as those that did not converge are not used in computing the histograms.}
\label{fig:DataHMSweep}
\end{figure*}

Extending the Gaussian task of Fig.~\ref{fig:Gaussian}c, we construct ID and OOD sets as fan-shaped regions in the input domain, such that the ID/OOD split can be varied systematically (Fig.~\ref{fig:theta}a). The ID and OOD datasets for the Gaussian task $\mathcal{D}_{\text{ID}}$, $\mathcal{D}_{\text{OOD}}$ are defined as follows:
\begin{subequations}
\begin{align}
\mathcal{D}_{\text{full}}&\equiv\left\{(x_{1},x_{2})|x_{1}^{2}+x_{2}^{2}\leq1\right\},\\
\mathcal{D}_{\theta}&\equiv\left\{(x_{1},x_{2})\left|\tan\theta>\left|\frac{x_{2}-x_{1}}{x_{2}+x_{1}}\right|\right.\right\},\\
\mathcal{D}_{\text{ID}}&\equiv\mathcal{D}_{\text{full}}\cap\mathcal{D}_{\theta}, \ \  \ 
\mathcal{D}_{\text{OOD}} \equiv\mathcal{D}_{\text{full}}-\mathcal{D}_{\text{ID}}.
\end{align}
\end{subequations}
For $\theta=0$, there is no ID data; at $\theta=\pi/4$, exactly two quadrants are in the ID dataset; all data becomes ID when $\theta =\pi/2$ (Fig.~\ref{fig:theta}b). We quantify the terminal OOD loss for the three architectures as a function of $\theta$ by calculating the trimmed mean (25 to 75 percentile) OOD mean squared error loss over 20 different instantiations of each network, as we vary $\theta$. Comparing the trimmed mean losses, we see that the modularised network converges to the correct 2D Gaussian solution by $\theta\geq\pi/4$, while the other two networks exhibit a much more gradual convergence to the solution, only arriving near it at $\theta=\pi/2$, where there are no OOD regions in the data (Fig.~\ref{fig:theta}c).
\\

We used the trimmed mean because in rare cases, the modularised network finds an alternative solution that has a loss much bigger than the mode of the distribution. For $\theta=\pi/4$, the training dataset $\mathcal{D}_{\text{ID}}$ specifies that the learned function $\hat{f}(x,y)$ satisfies
\begin{equation}
\hat{f}(x_{1},x_{2})=\exp\left[-\frac{1}{2\sigma^{2}}(x_{1}^{2}+x_{2}^{2})\right]\ \text{if }x_{1}x_{2}>0.
\label{eq:target}
\end{equation}
The natural guess is that $\hat{f}(x_{1},x_{2})$ is a 2D Gaussian in the whole $x_{1}$, $x_{2}$ plane, which can be written in the form
\eqS
\hat{f}(x_{1},x_{2})=\chi[\psi_{1}(x_{1})+\psi_{2}(x_{2})],
\eqE
where the single-input functions $\chi(x)$, $\psi_{1}(x)$, and $\psi_{2}(x)$ are 
\begin{subequations}
\begin{align}
\chi(x)&=e^{-\alpha x/(2\sigma^{2})},\\
\psi_{1}(x)&=x^{2}/\alpha,\\
\psi_{2}(x)&=x^{2}/\alpha,
\end{align}
\end{subequations}
for a gauge variable $\alpha$ and up to other additive gauge variables. In some rare cases, the molularised neural network however found an alternative solution:
\begin{subequations}
\begin{align}
\bar{\chi}(x)&=e^{-|x|\cdot\left[\beta+\alpha\Theta(x)\right]/(2\sigma^{2})},\\
\bar{\psi}_{1}(x)&=\text{sign}(x)\cdot x^{2}/\left[\beta+\alpha\Theta(x)\right],\\
\bar{\psi}_{2}(x)&=\text{sign}(x)\cdot x^{2}/\left[\beta+\alpha\Theta(x)\right],
\end{align}
\label{eq:altersol}
\end{subequations}
for an additional gauge variable $\beta$, and $\Theta(x)$ is the Heaviside step-function. For this alternative solution
\begin{equation}
\hat{\bar{f}}(x_{1},x_{2})=\bar{\chi}[\bar{\psi}_{1}(x_{1})+\bar{\psi}_{2}(x_{2})],
\label{eq:alterOverall}
\end{equation}
if $x_{1}$ and $x_{2}$ are of the same sign, we recover the 2-dimensional Gaussian. However, in the OOD quadrants, we take $x_{1}>0$ and $x_{2}<0$ without loss of generality, yielding
\begin{subequations}
    \begin{align}
        \bar{\psi}_{1}(x_{1})&=x_{1}^{2}/(\beta+\alpha),\\
        \bar{\psi}_{2}(x_{2})&=-x_{2}^{2}/\beta,
    \end{align}
\end{subequations}
and a `ridge' appears at a set of points $(x_{1}^{*},x_{2}^{*})$ when $\bar{\psi}_{1}(x_{1}^{*})+\bar{\psi}_{2}(x_{2}^{*})=0$ (Fig.~\ref{fig:theta}d):
\eqS
\frac{|x_{1}|}{|x_{2}|}=\sqrt{1+\frac{\alpha}{\beta}}.
\eqE
The alternative solution defined by Eqs.~\eqref{eq:altersol} is equally valid as far as the training data is concerned, and reflects a major difficulty in studying artificial neural network (ANN) generalisation: there is no `ground truth' per se, so it is difficult to determine whether an ANN learned to generalise. We hypothesise that the modularised network structure appears to have learned to generalise because the solution space for Eqs.~\eqref{eq:altersol} is a manifold disjoint from the 2D Gaussian in the function space, and which solution the network eventually finds depends on the initialisation as well as the training algorithm. In Fig.~\ref{fig:Gaussian}f, when computing the average over 20 different seeds for each $\theta$, around 2 and 3 initialisations out of 20 would find this alternative solution (Fig.~\ref{fig:theta}e). These have high OOD loss that deviate very much from the majority of the networks that converge to the Gaussian, so we trim the highest and lowest 5 OOD losses when computing the average over seeds.
\\

\balancecolsandclearpage

\bibliographystyle{ieeetr}

\begin{thebibliography}{10}

\bibitem{lorenzEmergenceModularityBiological2011}
D.~M. Lorenz, A.~Jeng, and M.~W. Deem, ``The emergence of modularity in
  biological systems,'' {\em Physics of Life Reviews}, p.~S1571064511000170,
  Feb. 2011.

\bibitem{chomskyASPECTSTHEORYSYNTAX}
N.~Chomsky, ``{{ASPECTS OF THE THEORY OF SYNTAX}},''

\bibitem{boopathyBREAKINGNEURALNETWORK2025}
A.~Boopathy, S.~Jiang, W.~Yue, J.~Hwang, A.~Iyer, and I.~Fiete, ``{{BREAKING
  NEURAL NETWORK SCALING LAWS WITH MODULARITY}},'' 2025.

\bibitem{veniatEFFICIENTCONTINUALLEARNING2021}
T.~Veniat, L.~Denoyer, and M.~Ranzato, ``{{EFFICIENT CONTINUAL LEARNING WITH
  MODULAR NETWORKS AND TASK-DRIVEN PRIORS}},'' 2021.

\bibitem{aletNeuralRelationalInference}
F.~Alet, E.~Weng, T.~{Lozano-P{\'e}rez}, and L.~P. Kaelbling, ``Neural
  {{Relational Inference}} with {{Fast Modular Meta-learning}},''

\bibitem{parascandoloLearningIndependentCausal2018}
G.~Parascandolo, N.~Kilbertus, M.~{Rojas-Carulla}, and B.~Sch{\"o}lkopf,
  ``Learning {{Independent Causal Mechanisms}},'' Sept. 2018.

\bibitem{ellefsenNeuralModularityHelps2015}
K.~O. Ellefsen, J.-B. Mouret, and J.~Clune, ``Neural {{Modularity Helps
  Organisms Evolve}} to {{Learn New Skills}} without {{Forgetting Old
  Skills}},'' {\em PLOS Computational Biology}, vol.~11, p.~e1004128, Apr.
  2015.

\bibitem{hamidiModularGrowthHierarchical2024}
M.~Hamidi, S.~Khajehabdollahi, E.~Giannakakis, T.~Sch{\"a}fer, A.~Levina, and
  C.~M. Wu, ``Modular {{Growth}} of {{Hierarchical Networks}}: {{Efficient}},
  {{General}}, and {{Robust Curriculum Learning}},'' June 2024.

\bibitem{mccourtNoisyDynamicalSystems2023}
T.~McCourt, I.~R. Fiete, and I.~L. Chuang, ``Noisy dynamical systems evolve
  error correcting codes and modularity,'' Apr. 2023.

\bibitem{guEmergenceReconfigurationModular2024}
S.~Gu, M.~G. Mattar, H.~Tang, and G.~Pan, ``Emergence and reconfiguration of
  modular structure for artificial neural networks during continual familiarity
  detection,'' {\em Science Advances}, vol.~10, p.~eadm8430, July 2024.

\bibitem{monteroROLEDISENTANGLEMENTGENERALISATION2021}
M.~L. Montero, C.~J.~H. Ludwig, R.~P. Costa, G.~Malhotra, and J.~S. Bowers,
  ``{{THE ROLE OF DISENTANGLEMENT IN GENERALISATION}},'' 2021.

\bibitem{liangCompositionalGeneralizationRequires2025}
Q.~Liang, D.~Qian, L.~Ziyin, and I.~Fiete, ``Compositional {{Generalization
  Requires More Than Disentangled Representations}},'' Jan. 2025.

\bibitem{liangHowDiffusionModels2024}
Q.~Liang, Z.~Liu, M.~Ostrow, and I.~Fiete, ``How {{Diffusion Models Learn}} to
  {{Factorize}} and {{Compose}},'' Oct. 2024.

\bibitem{whittingtonDisentanglementBiologicalConstraints2023}
J.~C.~R. Whittington, W.~Dorrell, S.~Ganguli, and T.~E.~J. Behrens,
  ``Disentanglement with {{Biological Constraints}}: {{A Theory}} of
  {{Functional Cell Types}},'' Mar. 2023.

\bibitem{camutoExplicitRegularisationGaussian}
A.~Camuto and M.~Willetts, ``Explicit {{Regularisation}} in {{Gaussian Noise
  Injections}},''

\bibitem{pooleAnalyzingNoiseAutoencoders2014}
B.~Poole, J.~{Sohl-Dickstein}, and S.~Ganguli, ``Analyzing noise in
  autoencoders and deep networks,'' June 2014.

\bibitem{softkyHighlyIrregularFiring1993}
W.~Softky and C.~Koch, ``The highly irregular firing of cortical cells is
  inconsistent with temporal integration of random {{EPSPs}},'' {\em The
  Journal of Neuroscience}, vol.~13, pp.~334--350, Jan. 1993.

\bibitem{gurResponseVariabilityNeurons1997}
M.~Gur, A.~Beylin, and D.~M. Snodderly, ``Response {{Variability}} of
  {{Neurons}} in {{Primary Visual Cortex}} ({{V1}}) of {{Alert Monkeys}},''
  {\em The Journal of Neuroscience}, vol.~17, pp.~2914--2920, Apr. 1997.

\bibitem{chenDevelopmentModularityNeural2015}
M.~Chen and M.~W. Deem, ``Development of modularity in the neural activity of
  children's brains,'' {\em Physical Biology}, vol.~12, p.~016009, Jan. 2015.

\bibitem{dq219HttpsGithubcomDq2192025b}
{dq219}, ``{{https://github.com/dq219/NoiseMod}},'' July 2025.

\bibitem{luxburgTutorialSpectralClustering2007}
U.~von Luxburg, ``A {{Tutorial}} on {{Spectral Clustering}},'' Nov. 2007.

\bibitem{liaoSelfAssemblyBiologicallyPlausible2024}
Q.~Liao, L.~Ziyin, Y.~Gan, B.~Cheung, M.~Harnett, and T.~Poggio,
  ``Self-{{Assembly}} of a {{Biologically Plausible Learning Circuit}},'' Dec.
  2024.

\bibitem{bishopTrainingNoiseEquivalent1995}
C.~M. Bishop, ``Training with {{Noise}} is {{Equivalent}} to {{Tikhonov
  Regularization}},'' {\em Neural Computation}, vol.~7, pp.~108--116, Jan.
  1995.

\bibitem{Debao1993}
C.~Debao, ``{Degree of approximation by superpositions of a sigmoidal
  function},'' {\em Approximation Theory and its Applications}, vol.~9, no.~3,
  pp.~17--28, 1993.

\bibitem{Zhou2020c}
D.~X. Zhou, ``{Universality of deep convolutional neural networks},'' {\em
  Applied and Computational Harmonic Analysis}, vol.~48, no.~2, pp.~787--794,
  2020.

\bibitem{Pinkus1999}
A.~Pinkus, ``{Approximation theory of the MLP model in neural networks},'' {\em
  Acta Numerica}, vol.~8, pp.~143--195, 1999.

\bibitem{Augustine2024}
M.~T. Augustine, ``{A Survey on Universal Approximation Theorems},'' no.~1,
  pp.~1--10, 2024.

\bibitem{Leshno1993}
M.~Leshno, V.~Y. Lin, A.~Pinkus, and S.~Schocken, ``{Multilayer feedforward
  networks with a nonpolynomial activation function can approximate any
  function},'' {\em Neural Networks}, vol.~6, no.~6, pp.~861--867, 1993.

\bibitem{williamsSimpleStatisticalGradientfollowing}
R.~J. Williams, ``Simple statistical gradient-following algorithms for
  connectionist reinforcement learning,''

\bibitem{seungLearningSpikingNeural2003}
H.~Seung, ``Learning in {{Spiking Neural Networks}} by {{Reinforcement}} of
  {{Stochastic Synaptic Transmission}},'' {\em Neuron}, vol.~40,
  pp.~1063--1073, Dec. 2003.

\bibitem{fieteGradientLearningSpiking2006}
I.~R. Fiete and H.~S. Seung, ``Gradient {{Learning}} in {{Spiking Neural
  Networks}} by {{Dynamic Perturbation}} of {{Conductances}},'' {\em Physical
  Review Letters}, vol.~97, p.~048104, July 2006.

\bibitem{srivastavaDropoutSimpleWay}
N.~Srivastava, G.~Hinton, A.~Krizhevsky, I.~Sutskever, and R.~Salakhutdinov,
  ``Dropout: {{A Simple Way}} to {{Prevent Neural Networks}} from
  {{Overfitting}},''

\bibitem{faisalNoiseNervousSystem2008}
A.~A. Faisal, L.~P.~J. Selen, and D.~M. Wolpert, ``Noise in the nervous
  system,'' {\em Nature Reviews Neuroscience}, vol.~9, pp.~292--303, Apr. 2008.

\bibitem{luRevealingDetailVisual2018}
Y.~Lu, J.~Yin, Z.~Chen, H.~Gong, Y.~Liu, L.~Qian, X.~Li, R.~Liu, I.~M.
  Andolina, and W.~Wang, ``Revealing {{Detail}} along the {{Visual Hierarchy}}:
  {{Neural Clustering Preserves Acuity}} from {{V1}} to {{V4}},'' {\em Neuron},
  vol.~98, pp.~417--428.e3, Apr. 2018.

\bibitem{hulseMechanismsUnderlyingNeural2020}
B.~K. Hulse and V.~Jayaraman, ``Mechanisms {{Underlying}} the {{Neural
  Computation}} of {{Head Direction}},'' {\em Annual Review of Neuroscience},
  vol.~43, pp.~31--54, July 2020.

\bibitem{lyuBuildingAllocentricTravelling2022}
C.~Lyu, L.~F. Abbott, and G.~Maimon, ``Building an allocentric travelling
  direction signal via vector computation,'' {\em Nature}, vol.~601,
  pp.~92--97, Jan. 2022.

\bibitem{haftingMicrostructureSpatialMap2005}
T.~Hafting, M.~Fyhn, S.~Molden, M.-B. Moser, and E.~I. Moser, ``Microstructure
  of a spatial map in the entorhinal cortex,'' {\em Nature}, vol.~436,
  pp.~801--806, Aug. 2005.

\bibitem{burakAccuratePathIntegration2009}
Y.~Burak and I.~R. Fiete, ``Accurate {{Path Integration}} in {{Continuous
  Attractor Network Models}} of {{Grid Cells}},'' {\em PLoS Computational
  Biology}, vol.~5, p.~e1000291, Feb. 2009.

\bibitem{kolmogorovRepresentationContinuousFunctions1963}
A.~N. Kolmogorov, {\em {On the representation of continuous functions of many
  variables by superposition of continuous functions of one variable and
  addition}}, vol.~28, pp.~55--59.
\newblock Providence, Rhode Island: American Mathematical Society, 1963.

\bibitem{schmidt-hieberKolmogorovArnoldRepresentation2021}
J.~{Schmidt-Hieber}, ``The {{Kolmogorov}}--{{Arnold}} representation theorem
  revisited,'' {\em Neural Networks}, vol.~137, pp.~119--126, May 2021.

\bibitem{duganOccamNetFastNeural2023}
O.~Dugan, R.~Dangovski, A.~Costa, S.~Kim, P.~Goyal, J.~Jacobson, and
  M.~Solja{\v c}i{\'c}, ``{{OccamNet}}: {{A Fast Neural Model}} for {{Symbolic
  Regression}} at {{Scale}},'' Nov. 2023.

\bibitem{liuKANKolmogorovArnoldNetworks2025}
Z.~Liu, Y.~Wang, S.~Vaidya, F.~Ruehle, J.~Halverson, M.~Solja{\v c}i{\'c},
  T.~Y. Hou, and M.~Tegmark, ``{{KAN}}: {{Kolmogorov-Arnold Networks}},'' Feb.
  2025.

\bibitem{Cranmer2020}
M.~Cranmer, A.~Sanchez-Gonzalez, P.~Battaglia, R.~Xu, K.~Cranmer, D.~Spergel,
  and S.~Ho, ``{Discovering symbolic models from deep learning with inductive
  biases},'' {\em Advances in Neural Information Processing Systems},
  vol.~2020-Decem, no.~NeurIPS, pp.~1--25, 2020.

\end{thebibliography}

\end{document}